\documentclass[aps,prd,superscriptaddress,showpacs,preprintnumbers]{revtex4}

\usepackage{color}
\usepackage{epsfig}
\usepackage{graphicx,graphics}
\usepackage{float}
\usepackage{amsmath}
\usepackage{amsfonts}
\usepackage{amssymb}
\usepackage{multirow}

\newcommand{\bi}{\begin{itemize}}
\newcommand{\ei}{\end{itemize}}

\usepackage{bm}

\newcommand{\nl}{\newline}
\newcommand{\beq}{\begin{equation}}
 \newcommand{\eeq}{\end{equation}}

 \newcommand{\be}{\begin{eqnarray}}
 \newcommand{\ee}{\end{eqnarray}}
 \newcommand{\nee}{\nonumber\end{eqnarray}}
 \newcommand{\nn}{\nonumber\\}

  \newcommand{\bc}{\begin{center}}
 \newcommand{\ec}{\end{center}}

\newcommand{\qup}{q^\uparrow}

\def\kt{k_\perp}

\def\pp{p_\perp}

\def\pt{P_T}
\def\avk{\langle k_\perp ^2\rangle}
\def\avp{\langle p_\perp ^2\rangle}
\def\avPT{\langle P_T^2\rangle}

\def\S{_{_S}}

\def\C{_{_C}}

\def\BM{_{_{B\!M}}}

\def\xb{x_{_{\!B}}}
\def\1{_{_{C\!1}}}
\def\2{_{_{S1}}}

\def\s              {\sigma}
\def\g              {\gamma}

\begin{document}

\author{E. Christova}
\email {echristo@inrne.bas.bg} \affiliation{Institute for Nuclear Research and Nuclear Energy, Bulgarian Academy of Sciences,
Tzarigradsko chauss\'{e}e 72, 1784 Sofia, Bulgaria}
\author{D. Kotlorz}
\email {dorota@theor.jinr.ru} \affiliation{Department of Physics, Opole University of Technology, 45-758 Opole, Proszkowska 76,
Poland} \affiliation{Bogoliubov Laboratory of Theoretical Physics, JINR, 141980 Dubna, Russia}
\author{E. Leader}
\email {e.leader@imperial.ac.uk} \affiliation{ Imperial College London, London SW7 2AZ, United Kingdom}
\title{New study of the Boer-Mulders function: Implications for the quark and hadron transverse momenta}

\begin{abstract}
In series of papers the Boer-Mulders function for a given quark flavour has been extracted: (i) from data on semi-inclusive
deep inelastic scattering, using  the simplifying, but theoretically inconsistent, assumption that it is proportional to
the Sivers function for each quark flavour and (ii) from data on Drell-Yan reactions. In earlier papers, using the
semi-inclusive deep inelastic COMPASS deuteron data on the $\langle\cos\phi_h\rangle$ and $\langle\cos 2 \phi_h\rangle$
asymmetries, we extracted the collinear $x_{_{\!B}}$ dependence of the Boer-Mulders function for the sum of the valence quarks
$Q_V=u_V + d_V$ using a small number of model dependent assumptions and found a significant disagreement with the analysis in (i). \nl
In the present paper, we provide a more complete analysis of the semi-inclusive deep inelastic scattering reaction, including a
discussion of higher twist and interaction-dependent terms, and also a comparison with the Boer-Mulders function extracted from
data on the Drell-Yan reaction. We confirm that the proportionality relation of the BM function to the Sivers function, for each
quark flavour, fails badly, but find that it holds rather well if applied to the nonsinglet valence-quark combination, $Q_V$.
We also find good agreement with the results of the Drell-Yan analysis. Furthermore, we obtain interesting information on
the quark transverse momentum densities in the nucleon and on the hadron transverse momentum dependence in quark fragmentation.
\end{abstract}

\date{\today}

\maketitle

\section{Introduction}
The Boer-Mulders (BM) function \cite{Boer:1997nt} is an essential element in describing the internal structure of the nucleon.
A nonzero BM function implies that inside an unpolarized  nucleon there are transversely polarized quarks. It is a leading
twist, chiral odd, transverse momentum dependent (TMD) parton distribution. In a nucleon of momentum $\bf P$, and for a quark with
transverse momentum ${\bf k}_\perp$, the BM function measures the difference between the number density of quarks polarized
parallel and antiparallel to $ {\bf ( P \times k_\perp)}$. It describes  the distribution of transversely polarized  quarks
$\qup$ in an unpolarized proton $p$. Different notations for it are found in literature,
 \be
 \Delta^N \! f_{\qup/p}(x_B, \kt) \equiv \Delta f^{q}_{s_y/p}(x_B,\kt)  = - \frac
{\kt}{m_p} \, h_{1}^\perp (x_B, \kt)\, . \label{b-m} \ee

First attempts to extract it from experiment were hindered by the scarcity of data and  made the  simplifying model assumption
of its proportionality to the better known Sivers function \cite{Barone:2008tn, Barone:2009hw, Barone:2010gk}, an assumption
motivated by model calculations \cite{Bacchetta:2008af, Courtoy:2009pc, Pasquini:2010af}. However,  this assumption applied for
each quark separately, as explained   in \cite{Christova:2017zxa, Christova:2019fbj}, is theoretically inconsistent as it leads
to gluons contributing in the evolution of nonsinglet combinations of quark densities.
Other parametrizations for the BM function were obtained in \cite{Zhang:2008nu, Lu:2009ip} from
 data on the $pD$ and $pp$ Drell-Yan (DY) processes. These processes are controlled by products of  two
 BM functions
 $h^{q\perp}_1 (x_1, k_{\perp 1} )\,\cdot\,h^{q\perp}_1 (x_2, k_{\perp 2})$, one from each of the initial hadrons
in the reaction, and an additional condition, the positivity bound, is
used to constrain some of the parameters. In \cite{Wang:2018naw}, the transverse momentum dependent evolution of the
pion and proton BM functions  was studied in the pion induced DY process $\pi +N \rightarrow {l^+}{l^{-}} + X$.

In this paper, we show that the new COMPASS data on the unpolarized $\langle\cos\phi_h\rangle$ and $\langle\cos 2\phi_h\rangle$
asymmetries in semi-inclusive deep inelastic scattering (SIDIS) reactions for producing a hadron $h$ and its antiparticle $\bar
h$ at azimuthal angle $\phi_h$ allow an extraction of the BM function with a  small number of model-dependent assumptions.

\section{The Formalism}

As explained in \cite{Christova:2000nz} and \cite{Christova:2015jsa}, there is a great advantage in studying difference
asymmetries $A^{h-\bar h}$, effectively $ A^h -A^{\bar h}$, since both for the collinear and TMD
functions, only the flavour nonsinglet  valence quark parton densities
[parton distribution functions (PDFs)] and fragmentation functions (FFs) play a role and
the gluon does not contribute. On a deuteron target, an additional simplification occurs. Independently of the final hadron,
only the sum of the valence-quark TMD functions  $Q_V=u_V+d_V$ enters. The above statements are general, based only
on factorization  of the scattering and fragmentation processes in SIDIS, and on  the C and SU(2) invariance of strong
interactions, with no assumptions on the parametrizations of the TMD-PDFs and TMD-FFs (see \cite{Christova:2014gva}).

In this paper, we apply the method of the difference asymmetries to the latest SIDIS COMPASS data \cite{Adolph:2014pwc} on a
deuteron target,  aiming to extract the BM TMD function for $Q_V$ with a small number of model-dependent assumptions. The first
step to achieve this program is to choose a definite parametrization for the Boer-Mulders function.
As is  often done, we assume  factorized $\xb$- and $k_\perp$-dependent functions, each  proportional to the  relevant
unpolarized TMD function.
Thus, for the BM function for $Q_V$, we adopt a Gaussian distribution for the $k_\perp$ dependence with  
 $\xb$-independent width, and the collinear $\xb$-dependent part, is obtained  by multiplying the
unpolarized distribution by a fitted flavour-dependent function of $\xb$.  Note that the question of flavour-dependent
widths for the individual quark contributions to the BM function does not arise, because in this paper we need only,
and parametrize only, the contribution of the combination $Q_V$. The simplified parametrization  has been
often used in the literature.

  Recently, in a number of papers \cite{Bacchetta:2017gcc, Bacchetta:2019sam, Scimemi:2017etj, Scimemi:2019cmh},
  the unpolarized TMDs, which are basic in all TMD analysis, have been
  extracted in  global analyses of  multiplicities in SIDIS, of DY reactions and of $Z$ production. As these processes are at
  quite different $Q^2$ ranges--from
  $Q \sim$  few GeV in SIDIS  up to $Q\sim M_Z$ or larger in $Z$ production, TMD evolution is necessarily applied, and in
  addition more general forms of $\kt$ dependencies have been tested--with flavour or $Q^2$-dependent Gaussian widths etc.
  \cite{Bacchetta:2017gcc, Bacchetta:2019sam, Scimemi:2017etj, Scimemi:2019cmh, Signori:2013mda, Anselmino:2012aa, Bertone:2019nxa}.
  However, the precise form of the TMD evolution is controversial, as will be explained in
  Sec. V, and
      the shortage  of  data on the  azimuthal asymmetries considered in this paper, and  their lack of precision
      suggest that it would be impossible to make any  meaningful assessment of these refinements  for the BM
      function. Moreover, as will be shown, we achieve an excellent fit to the data using our simplified forms without
      evolution. In addition, one of our goals is to compare our extracted BM function to  the existing parametrizations
      in the literature, \cite{Barone:2009hw, Barone:2010gk, Zhang:2008nu, Lu:2009ip},
      in which analogous simplifying assumptions have been  made.

\subsection{Parametrization of the TMD functions}

The  unpolarized TMD functions for $Q_V$ are parametrized in the often used
simplified  form~\cite{Christova:2014gva, Anselmino:2011ch} as a product of a function of $(x_B, Q^2)$ and a function of
$k_\perp^2$ or $p_\perp^2$. But although such a factorization, strictly speaking, is impossible,  it is perfectly acceptable to
use it to  provide a  parametrization of data  in some limited kinematic range. For studies where the
kinematic range is much greater than in SIDIS reactions, for example, in Drell-Yan and $Z$ production,
more general functional forms have been explored. See, for example, \cite{Bacchetta:2017gcc, Scimemi:2017etj, Signori:2013mda}.
We take

\be f_{Q_V/p}(\xb,k_\perp^2 , Q^2)&=&Q_V(\xb,Q^2)\,\frac{e^{-k_\perp^2/\avk}}{\pi\avk}\label{fq} \ee
 and
 \be
D_{h/q_V}(z_h,p_\perp^2 ,Q^2) =D_{q_V}^h(z_h,Q^2) \,\frac{e^{-p_\perp^2/\avp }}{\pi\avp },\label{Dq} \ee where $Q_V(\xb,Q^2 )$
is the sum of the collinear valence-quark  PDFs,
 \be
 Q_V(\xb,Q^2)= u_V(\xb,Q^2) +d_V(\xb,Q^2)
  \ee
  and $D_{q_V}^h(z_h,Q^2)$ are the valence-quark collinear  FFs,
   \be
  D_{q_V}^h(z_h,Q^2)=D_{q}^h(z_h,Q^2)-D_{\bar q}^h(z_h,Q^2),
  \ee
  and  $\avk $ and $\avp$ are
parameters extracted from a study of the multiplicities in unpolarized SIDIS.
There is some controversy in the literature, with several different published sets of values. It will turn out that this
analysis favours a particular choice  of these values.

The BM function is parametrized in a similar way, \beq \hspace*{-.5cm}  \Delta  f^{Q_V}\BM (\xb,\kt ,Q^2) \!=\! \Delta
f^{Q_V}\BM (\xb,Q^2)\; \sqrt{2e}\,\frac{\kt}{M\BM } \; \frac{e^{-\kt^2/\avk\BM  }}{\pi\avk },\label{BM-Siv_dist1} \eeq with \beq
\Delta f^{Q_V}\BM (\xb,Q^2)\!=\! 2\,{\cal N}\BM ^{Q_V}(\xb)\,Q_V(\xb,Q^2). \label{BM-Siv_dist2}
 \eeq
Here the ${\cal N}^{Q_V}\BM (\xb)$ is an unknown function and $M\BM $,  or equivalently $\avk\BM$, \be
 \avk \BM = \frac{\avk  \, M^2\BM }{\avk  + M^2 \BM },
 \ee
is an unknown parameter.

Since the asymmetries under study involve a product of the BM parton density and the Collins FF, one requires also the
transverse momentum dependent Collins function \cite{Collins:1992kk}, which is also parametrized in the often used simplified way,
\beq
 \Delta^N  D_{h/q_V\uparrow}(z_h,\pp ,Q^2) \!=\!\Delta^N  D_{h/q_V\uparrow}(z_h,Q^2)\,
\sqrt{2e}\,\frac{\pp}{M\C} \; \frac{e^{-\pp^2/\avp\C }}{\pi\avp}\,, \label{form9} \eeq where \beq \Delta^N
D_{h/q_V\uparrow}(z_h,Q^2)\!=\!2\,{\cal N}^{h/\!q_V}\C (z_h)\,D_{q_V}^h(z_h,Q^2).
\eeq The quantities  ${\cal N}^{h/\!q_V}\C (z_h)$ and $M\C$,
or equivalently $ \avp\C$, \be
 \avp\C  =\frac{\avp  \, M\C ^2}{\avp  +M\C ^2}
\label{Coll-frag2}
 \ee
are known from studies of the azimuthal correlations of pion-pion, pion-kaon and kaon-kaon pairs produced in $e^+e^-$
annihilation: $e^+e^-\to h_1h_2+X$ and the $\sin (\phi_h +\phi_S)$ asymmetry in polarized SIDIS \cite{Anselmino:2008jk,
Anselmino:2015sxa, Anselmino:2015fty}.

Besides the BM-Collins contributions to the $\langle\cos\phi_h\rangle$ and $\langle\cos 2\phi_h\rangle$ unpolarized asymmetries,
there exists also a contribution known as the Cahn effect \cite{Cahn:1978se, Cahn:1989yf}. The Cahn effect is a purely kinematic
effect, generated in the naive parton model by the quark intrinsic transverse momenta included in distribution and fragmentation
functions. It is described by the unpolarized TMD functions $f_1^q (\xb ,\kt )$ and $D_{1q}^h(z_h,\pp )$, and is a subleading
effect, i.e., $1/Q^2$ contribution to the asymmetry $A^{\cos 2 \phi}_{UU}$,
and a $1/Q$ contribution to $A^{\cos  \phi}_{UU}$.

It should be noted that our intention is to use Eqs. (\ref{fq}), (\ref{Dq}), (\ref{BM-Siv_dist1}) and (\ref{form9})
only at some average $Q^2$,  relevant for the available data.
Using these equations to evolve in $Q^2$ would be incorrect as is explained in Sec. V.

\subsection{The difference asymmetries}

 For the differential cross section for SIDIS of unpolarized leptons on unpolarized
nucleons  in the considered kinematic region $\pt \simeq k_\perp \ll Q$, we use the expression \cite{Anselmino:2011ch}
 \be
 \frac{d\s_N^{h}}{dx_B\,dQ^2\,dz_h\,d^2{\bf P}_T  }&=&
\s_{0}^{h}\left\{1+ \frac{2(1-y)}{[1+(1-y)^2]}\,\cos 2\phi_h\,A_{UU}^{\cos 2\phi_h ,h}+
\frac{2(2-y)\sqrt{1-y}}{[1+(1-y)^2]}\,\cos \phi_h\,A_{UU}^{\cos \phi_h,h}\right\}\label{1}
 \ee
 where $\s_0^h$ is the $\phi_h$-independent part of the cross section and, $A_{UU}^{\cos \phi_h,h}$ and $A_{UU}^{\cos 2\phi_h,h}$ are 
 the $\cos\phi_h$ and $\cos 2\phi_h$  azimuthal asymmetries  measured at COMPASS \cite{Adolph:2014pwc} that we shall consider.
 [Note that in our previous paper \cite{Christova:2017zxa},
 following the Trento convention \cite{Bacchetta:2006tn}, different  definitions for the asymmetries were used, in which
 the kinematic $y$-dependent prefactors were incorporated in the symbols 
   $A_{UU}^{\cos \phi_h,h}$ and $A_{UU}^{\cos 2\phi_h,h}$.] They are 
  generated by the two contributions--the Cahn and the Boer-Mulders TMD mechanisms.
The  $\cos\phi_h$ asymmetry  gets  twist-3 Cahn and BM contributions as well as interaction-dependent terms associated with
quark-gluon-quark correlators  \cite{Bacchetta:2006tn},
 which will be discussed later. The $\cos 2\phi_h$ term is generated by a leading twist-2 BM effect and a twist-4 Cahn effect.
 The twist-4 Cahn term is only a part of the not yet calculated overall twist-4 contribution to the $\langle \cos 2\phi_h\rangle$
 asymmetry, 
 like hadron-mass corrections, etc. However, as we shall argue in Sec.~V, the Cahn contribution
is particularly important in the $\cos 2\phi_h$ asymmetry and neglecting it, as in the analysis in \cite{Barone:2015ksa},
is not justified.

In the above,  $P_T$ and $\phi_h$ are  the transverse momentum and azimuthal angle of the final hadron  in the $\g^*$-nucleon\,
c.m. frame. $z_h$, $Q^2$, and $y$ are the usual measurable SIDIS quantities,
\be
 \quad z_h=\frac {(P\cdot P_h)}{(P\cdot q)},\quad Q^2=-q^2, \quad q=l-l',
\quad
 y=\frac{(P\cdot q)}{(P\cdot l)}
 \ee
 where $l$ and $l'$, $P$ and $P_h$  are the 4-momenta of the initial
and final leptons,  and the initial and final hadrons.  Note that
 \be
 \quad Q^2=2ME \xb y
\ee where $M$ is the target mass (in this paper the deuteron
  mass) and $E$ the lepton laboratory energy.

Further, we shall work with the so-called difference asymmetries \cite{Christova:2000nz, Christova:2014gva} that have the
following general structure:
  \be
  A^{h^+ - h^-}\equiv \frac{\Delta \s^{h^+}-\Delta \s^{ h^-}}{\s^{h^+} -\s^{ h^-}},\label{A}
\ee
  where $\s^{h^+, \,h^-}$   and $\Delta \s^{h^+, \,h^-}$ are
    the  unpolarized and polarized  cross sections respectively.
   The difference asymmetries are not a new measurement, but they are
 expressed in terms of the usual  asymmetries $ A^{h^+,\, h^-}$,
  \be
  A^{h^+}= \frac{\Delta \sigma^{h^+}}{\sigma^{h^+}}, \qquad A^{h^-}= \frac{\Delta \sigma^{h^-}}{\sigma^{h^-}},
   \ee
 and the ratio $r$ of the unpolarized  SIDIS
 cross sections for production of  $h^-$ and $h^+$,
 $ r = \sigma^{h^-} / \sigma^{h^+}$,
 \beq
    A^{h^+ -h^-} = \frac{1}{1-r} \left( A^{h^+} - r A^{h^-}\right) .
    \label{eq.Diff1}
     \eeq
  As mentioned above, the advantage of using the difference asymmetries is that,
   based only on charge conjugation
(C) and isospin (SU(2)) invariance of the strong interactions,
   they are expressed  purely in terms of the best known valence-quark
    distributions and  fragmentation functions;
 sea-quark and gluon distributions do not enter. For a deuteron target, there is the additional simplification that,
 independently of the final hadron,  only the sum of the valence-quark distributions   enters.  This simplifying feature, as has
 been mentioned above, is independent of the form of the parametrizations assumed for the various distributions and
 fragmentation functions. \\

 In the following, we use the  asymmetries $A_{UU}^{\cos  \phi_h}$ and $A_{UU}^{\cos  2\phi_h}$ as defined in Eq. (\ref{1}) and
 used in the COMPASS paper~\cite{Adolph:2014pwc}.
 (Note that several different definitions \cite{DAlesio:2007bjf} of these asymmetries exist in the literature, some of them
even differing between COMPASS publications \cite{Bradamante:2007ex}). Neglecting the $Q^2$ evolution  of the
collinear PDFs and FFs in the considered  kinematic range involved,
 the $\xb$-dependent difference asymmetries are related to the theoretical functions via
    \be
  A_{UU}^{\cos  \phi_h, h-\bar h}(\xb )&=&\sqrt \frac{\avk}{\langle Q^2 (\xb)\rangle }\,
  \left\{{\cal N}\BM^{Q_V}(\xb)\,{\cal C}\BM^h +{\cal C}_{Cahn}^h\right\},   \label{A1}\\
   A_{UU}^{\cos  2\phi_h, h-\bar h}(\xb )&=&\left\{{\cal N}\BM^{Q_V}(\xb)\,
   \hat {\cal C}\BM^h + \frac{\avk}{\langle Q^2 (\xb)\rangle }\,\hat {\cal C}^h_{Cahn}\right\},\label{A2}
    \ee
where $\langle Q^2 (\xb)\rangle $ is some mean value of $Q^2$ for the corresponding $\xb$ bin, and the coefficients ${\cal
C}\BM$, ${\cal C}_{Cahn}$, $\hat {\cal C}\BM$, and $\hat {\cal C}_{Cahn}$ are dimensionless constants given by integrals over
various products of  the unpolarized  or Collins  FFs and, crucially, whose values depend on  the parameters $\avk$, $\avp$,
$M\BM$, and $M\C$. For a finite range of integration over $P_T^2$, corresponding to the experimental kinematics, $a\leq P_T^2\leq
b$, they are given by the expressions \be {\cal C}_{Cahn}^h &=& -2\,\frac{\int dz_h\,z_h\,[D_{q_V}^{h}(z_h)]S_1(a,b;
\avPT)/(\eta + z_h^2)^{1/2}}{\int dz_h\, [D_{q_V}^{h}(z_h)]\,S_0(a,b; \avPT)} \label{Cahn} \\ {\cal C}\BM^h &=&
4e\,\frac{\lambda\BM^2\lambda\C^2}{M\BM M\C}\,\avp \nonumber \\ &&\times\frac{\int dz_h\,[\Delta^N
D_{{q_V}\!\uparrow}^{h}(z_h)]\,
[\,z_h^2\lambda\BM\,S_3(a,b,\avPT\BM)+\,(\eta\lambda\C-z_h^2\lambda\BM)\,S_1(a,b;\avPT\BM)]/(z_h^2\lambda\BM+\eta\lambda\C)^{3/2}}
{\int dz_h\, [D_{q_V}^{h}(z_h)]\,S_0(a,b; \avPT)}\label{BM}
\\
\hat {\cal C}^h_{Cahn} &=& \frac{2\int dz_h \left(z_h^2/[\eta +z_h^2]\right)\,[D_{q_V}^{h}(z_h)]\,S_2(a,b;\avPT)}{\int dz_h\,
[D_{q_V}^{h}(z_h)]\,S_0(a,b; \avPT)}\label{hatC}\\ \hat {\cal C}\BM^h &=& -2e\,\frac{\lambda\BM^2\lambda\C^2}{M\BM M\C}\,\avp\;
\frac{\int dz_h\,[z_h\,\Delta^N D_{{q_V}\!\uparrow}^{h}(z_h)]/(z_h^2\lambda\BM+\eta\lambda\C)\,S_2(a,b;\avPT\BM)} {\int dz_h\,
[D_{q_V}^{h}(z_h)]\,S_0(a,b; \avPT)}\label{hatBM} \ee where for $\tau = \textrm{either}\,\, \avPT\,\, \textrm{or}
\,\,\avPT\BM $, \be
 S_n(a,b; \tau)&=&\int_a^b\,
dP_T^2\,P_T^n e^{-P_T^2/\tau}/\tau^{1+n/2}\, .
 \ee
 \be
\avPT &=&\avp +z_h^2 \avk ,\qquad \avPT\BM =\avp\C +z_h^2 \avk\BM . \ee Here $[D_{q_V}^h (z_h)]$ and $ [\Delta^N
D_{{q_V}\!\uparrow}^{h}(z_h)]$ are combinations of the collinear and Collins FFs, \be [D_{q_V}^{h}(z_h)] &=&
e_u^2\,D_{u_V}^h+e_d^2\, D_{d_V}^h, \ee \be [\Delta^N D_{{q_V}\!\uparrow}^h (z_h)] &=& e_u^2\,\Delta^N D_{{u_V}\!\uparrow}^h
+e_d^2\,\Delta^N D_{{d_V}\!\uparrow}^h \label{D2} \ee
 and
   \be
    \eta =
\frac{\avp}{\avk},\quad \lambda\C  &=&\frac{ M\C ^2}{\avp +M\C ^2},\quad
 \lambda\BM  =\frac{ M\BM ^2}{\avk  +M\BM ^2}\cdot \label{CBM}
 \ee

\subsection{ The parameters $\langle k_\perp ^2\rangle$, $\langle p_\perp ^2\rangle$,
$M{_{_{B\!M}}^2}$, and $M{_{_C}^2}$}

As mentioned, there is  a wide range of values for these parameters given in the literature. The parameters  $\avk$ and $\avp$
are basic  as they enter the normalization functions in all TMD asymmetries. At present, the experimentally obtained values are
controversial:

 1. $\avk \approx 0.25\,\rm{GeV^2}$ and $\avp \approx 0.20\,\rm{GeV^2}$~\cite{Anselmino:2005nn}, extracted from the
  old EMC~\cite{Arneodo:1986cf} and FNAL~\cite{Adams:1993hs} data on the Cahn effect in the SIDIS $\cos\phi_h$ asymmetry.

 2. $\avk = 0.18\,\rm{GeV^2}$ and $\avp = 0.20\,\rm{GeV^2}$ \cite{Giordano:2008th}, based on a study of the old
  HERMES data on  the $\cos\phi_h$ and  $\cos 2\phi_h$ asymmetries in SIDIS.
  These values were used in the  extraction of the BM functions in \cite{Barone:2009hw}.

An analysis \cite{Anselmino:2013lza} of the more recent available data on multiplicities in SIDIS from
HERMES~\cite{Airapetian:2012ki} and COMPASS~\cite{Adolph:2013stb} separately gives quite different values:

3. $\avk = 0.57 \pm 0.08\,\rm{GeV^2}$ and $\avp = 0.12\pm 0.01\,\rm{GeV^2}$, extracted from  HERMES data

4. $\avk = 0.61 \pm 0.20\,\rm{GeV^2}$ and $\avp = 0.19\pm 0.02\,\rm{GeV^2}$, extracted from  COMPASS data.\\ Recently, the
importance of determining the values of $\avk$ and $\avp$ was specially stressed  \cite{Anselmino:2018psi}. Two quite different
parametrizations for both the Sivers \cite{Anselmino:2008sga, Anselmino:2016uie} and Collins \cite{Anselmino:2015fty,
Anselmino:2007fs} functions, with comparable accuracies of the fits to the data exist, but using  two very different values of
the Gaussian widths $\avk$ and $\avp$ of the unpolarized distributions.

We shall attempt to fit the SIDIS data using five different sets of the parameters $\avk$, $\avp$, $M\BM^2$, and $M\C^2$. For
$M_{BM}^2$, we try the values $0.34$, $0.19$, and $0.80 \, \rm{GeV^2} $, which correspond to the values for the Sivers $M_S^2$
obtained in \cite{Anselmino:2008sga} and \cite{Anselmino:2011gs, Alekseev:2008aa}. The value of $M\C^2$ is taken from the known
parametrizations of the Collins function \cite{Anselmino:2015sxa}  and \cite{Anselmino:2015fty}.

The coefficients ${\cal C}_{Cahn}$, ${\cal C}\BM$, $\hat {\cal C}_{Cahn}$, $\hat {\cal C}\BM$ are given in Table~I, grouped
together in sets corresponding to the values of these parameters,  with $\rho\equiv-{\cal C}\BM/\hat {\cal C}\BM$.\\

\begin{table}[h]
\begin{tabular}{|c|c|c|c|c|c|c|c|c|c|}
\hline Set & $~\avk~$ & $~\avp~ $ &$~M\BM^2~$  &$~M\C^2~$
 & ${\cal C}_{Cahn}$ & $~{\cal C}\BM~$ & $\hat {\cal C}_{Cahn}$   & $~\hat {\cal C}\BM~$ & $~~\rho~~$ \\ \hline
I    & 0.18 & 0.20 & 0.34 & 0.91 & $-0.68$ & 2.1 & 0.31 & $-0.47$ & 4.4 \\
II   & 0.18 & 0.20 & 0.19 & 0.91 & $-0.68$ & 1.8 & 0.31 & $-0.40$ & 4.4 \\
III  & 0.25 & 0.20 & 0.34 & 0.91 & $-0.77$ & 1.9 & 0.38 & $-0.49$ & 3.8 \\
IV   & 0.25 & 0.20 & 0.19 & 0.91 & $-0.77$ & 1.4 & 0.38 & $-0.39$ & 3.7 \\
V    & 0.57 & 0.12 & 0.80 & 0.28 & $-1.2$  & 0.89& 0.84 & $-0.50$ & 1.8 \\ \hline
\end{tabular}
\caption {${\cal C}_{Cahn}$, ${\cal C}\BM$, $\hat {\cal C}_{Cahn}$, $\hat {\cal C}\BM$, and $\rho$ calculated for different sets
of $\avk$, $\avp$, $M\BM^2$,  and $M\C^2$ $[{\rm GeV^2}]$. The parametrizations for the collinear FFs are from AKK'2008
\cite{Albino:2008fy}, and for the Collins functions--for sets I-IV--from \cite{Anselmino:2008jk} and
\cite{Anselmino:2015fty}, and for Set V--from \cite{Anselmino:2015sxa}  and \cite{Anselmino:2015fty}. The integrations  are
according to COMPASS kinematics:
 $0.01\leq P_T^2\leq 1 \,\rm{GeV^2}$ and $0.2\leq z_h\leq 0.85$ \cite{Adolph:2014pwc}.}
\end{table}

\section{The COMPASS asymmetries}

As mentioned earlier, we extract ${\cal N}\BM (\xb)$ from  the difference asymmetries $A^{h^+-h^-}_J$, related in
\cite{Christova:2017zxa} to the corresponding usual asymmetries $A_j^{h^+}$ and $A_j^{h^-}$ for positive  and negative charged
hadron production measured in COMPASS \cite{Adolph:2014pwc} via the relation~\cite{Alekseev:2007vi}
 \be
    A_J^{h^+-h^-} = \frac{1}{1-r} \left( A_J^{h^+} - r A_J^{h^-}\right),\qquad
J=\cos\phi_h,\; \cos 2\phi_h .
    \label{eq.Diff}
\ee Here $r$ is the ratio of the unpolarized $\xb$-dependent SIDIS cross sections for production of  negative and positive
hadrons $ r = \sigma^{h^-}(\xb ) / \sigma^{h^+}(\xb )$ measured in the same kinematics \cite{Alekseev:2007vi}. In the
COMPASS kinematics to each value of $\langle Q^2\rangle$ corresponds a definite value of $\langle \xb\rangle$:
thus, fixing the $Q^2$ interval, we fix also the $\xb$ interval. As shown in \cite{Christova:2017zxa}, in the whole
$Q^2$ range covered by COMPASS, $Q^2 \simeq [1-11]\,{\rm GeV^2}$, there is almost no $Q^2$ dependence both in the valence-quark
distributions $u_V$ and $d_V$ and in the FFs, i.e.,  in the whole $\xb$ interval. Thus, we consider it reasonable to use  our
simplified expressions (\ref{A1}, \ref{A2}) in the following interval $\xb\in[0.006,0.1]$ corresponding to
$Q^2\in[1.26,11.24]\,{\rm GeV^2}$.

\begin{figure}[H]
\begin{center}
\includegraphics[scale=0.5]{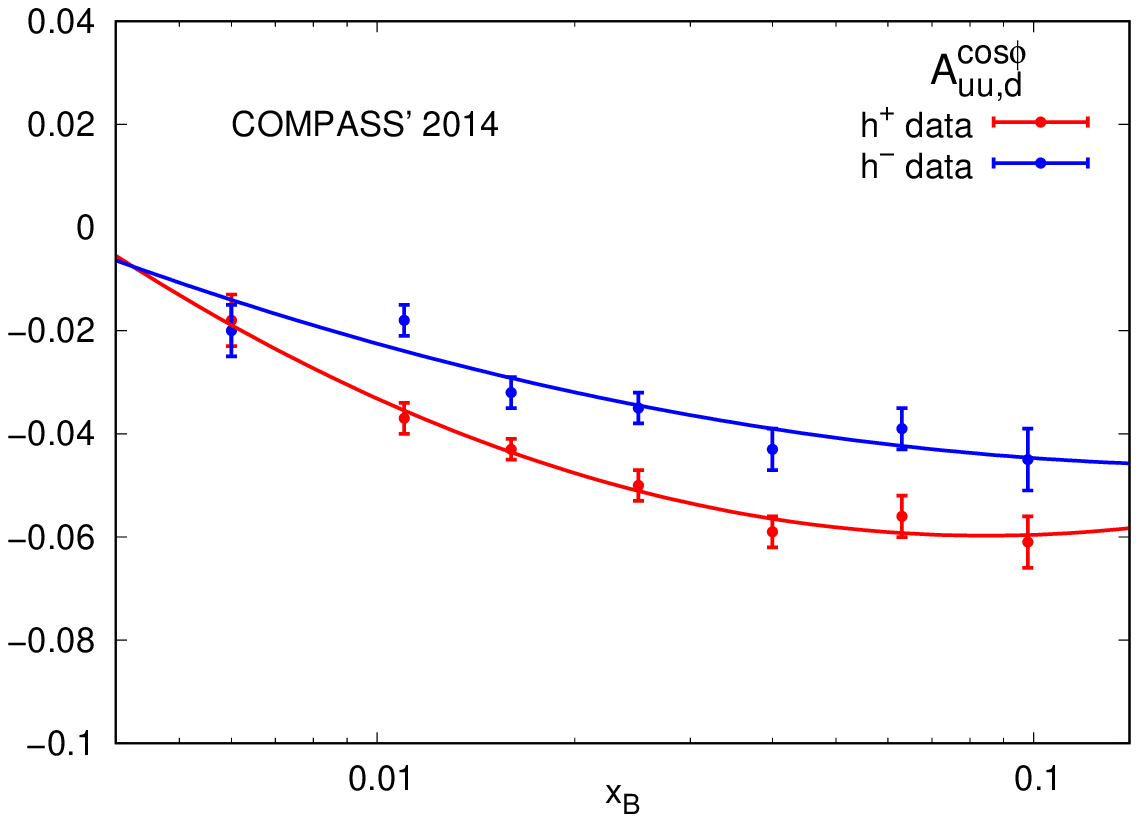}~~~\includegraphics[scale=0.5]{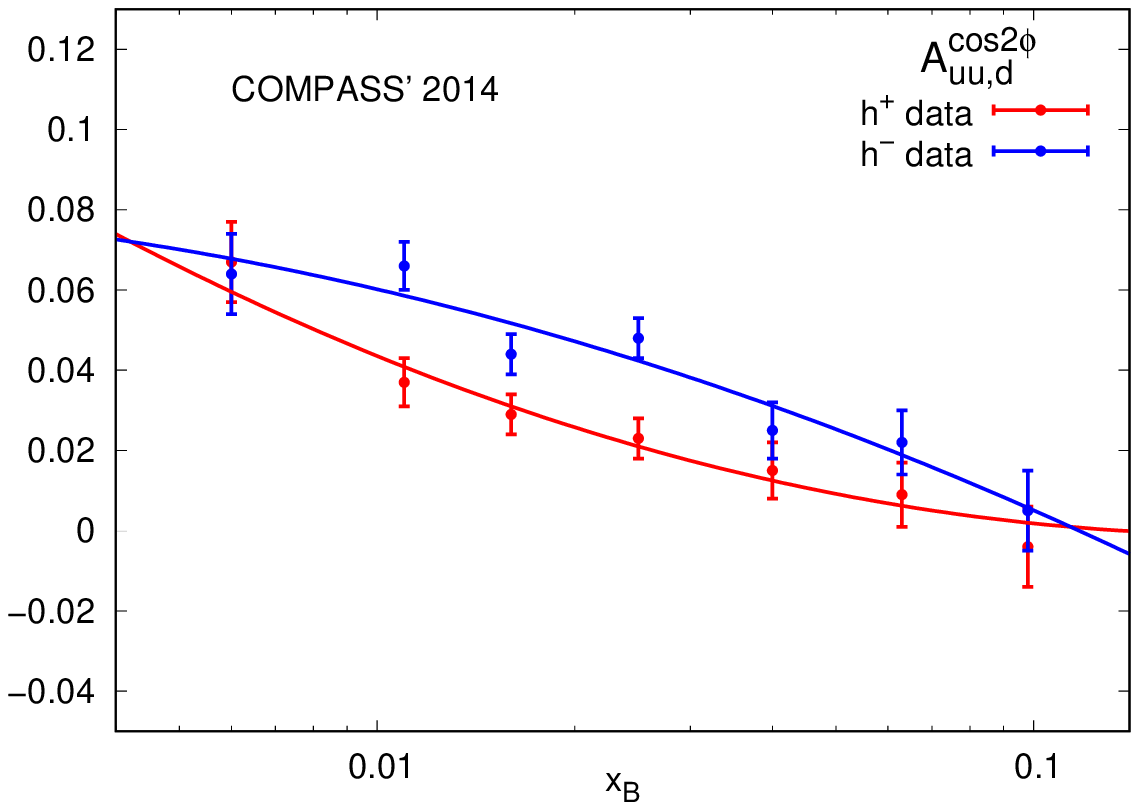}
\end{center}
\caption {Smoothed functional fit  to the ordinary asymmetries. $\chi^2_{d.o.f}=$ 0.49, 1.99, 0.47, and 1.56 for
$A_{UU,d}^{\cos\phi, h^+}$, $A_{UU,d}^{\cos\phi,h^-}$, $A_{UU,d}^{\cos 2\phi,h^+}$ and $A_{UU,d}^{\cos 2\phi,h^-}$,
respectively. Only statistical errors are shown.} \label{pic1}
\end{figure}

\begin{figure}[H]
\begin{center}
\includegraphics[scale=0.5]{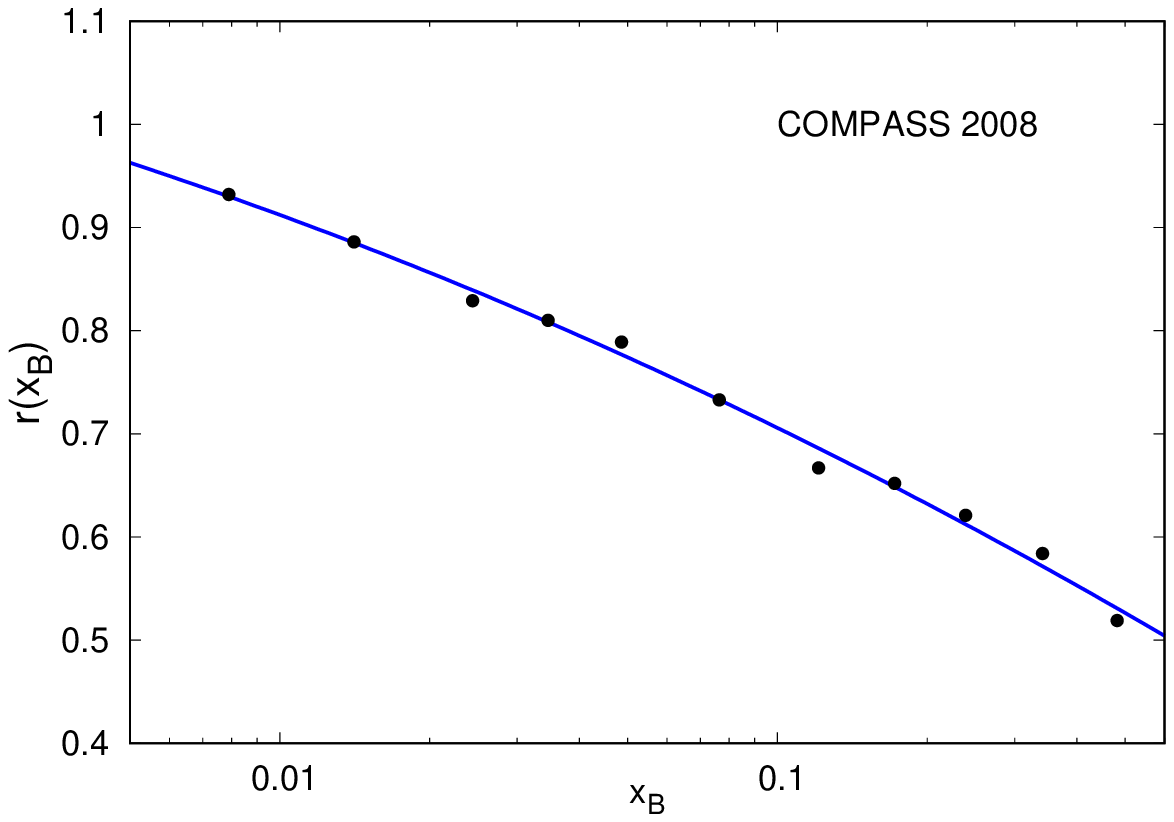}~~~\includegraphics[scale=0.5]{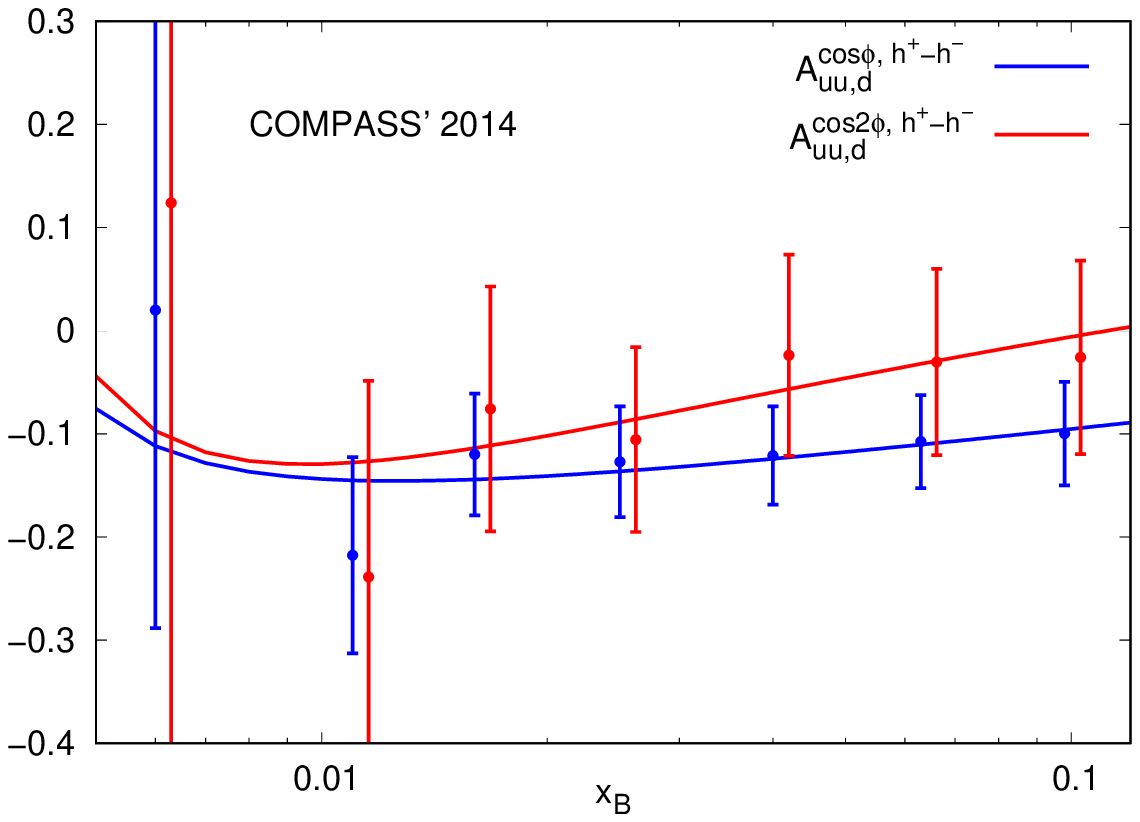}
\end{center}
\caption {Left: smoothed functional fit for $r$ from COMPASS data \cite{Alekseev:2007vi}. Right: for the difference
asymmetries, Eq.~(\ref{eq.Diff}). $\chi^2_{d.o.f}=$ 0.98 and 0.78 for $A_{UU,d}^{\cos\phi, h^+-h^-}$ and $A_{UU,d}^{\cos
2\phi,h^+-h^-}$, respectively. The points for $A_{UU,d}^{\cos 2\phi,h^+-h^-}$ are slightly shifted horizontally for better
visibility. Combined statistical and systematic experimental errors are shown.} \label{pic2}
\end{figure}

In our analysis, we use smooth fit functions of $\xb$ to the measured asymmetries $A_{UU}^{\cos  \phi,{h^{\pm}}}$, $A_{UU}^{\cos
2\phi,{h^{\pm}}}$, $r$, and $\langle Q^2(x_B)\rangle$. Then the difference asymmetries are calculated from Eq.~(\ref{eq.Diff}).
Our input functions are shown in Figs.~\ref{pic1} and \ref{pic2}. The error for the difference asymmetries is calculated as a
composed error implied by Eq.~(\ref{eq.Diff}), \beq \Delta A_{UU,d}^{j, h^+-h^-} = \frac{1}{1-r}\sqrt{(\Delta A_{UU,d}^{j,
h^+})^2 + r^2 (\Delta A_{UU,d}^{j, h^-})^2}\, ,\label{diff}
 \eeq
where $(\Delta \,A^{j})$ are the errors of the parameters in the function used to fit the asymmetries.
In the analysis, both statistical and systematic experimental errors are included.


\section{Numerical results on the BM function, $\langle k_\perp ^2\rangle$ and $\langle p_\perp ^2\rangle$}

Here, we present the strategy of our analysis and the obtained results.

\subsection{Compatibility extraction of the Boer-Mulders function}

We extract ${\cal N}\BM (\xb)$ from  relations  (\ref{A1}) and (\ref{A2}) of the difference asymmetries . Relations  (\ref{A1})
and (\ref{A2}) provide two independent equations for the extraction of ${\cal N}\BM (\xb)$ for each set of the parameters in
Table~I. The analysis shows that the two extractions are compatible with each other, within  errors, for the parameter values
$\{\avk,\; \avp\}$, \be &&\avk =0.18\;{\rm GeV^2},\qquad\avp =0.20\;{\rm GeV^2}\\ &&\avk =0.25\;{\rm GeV^2},\qquad\avp
=0.20\;{\rm GeV^2}, \label{best} \ee
  with a slight preference for (\ref{best}). Note that these  values for $\avk$ and $\avp$ agree with those obtained in
  \cite{Giordano:2008th}
and with the theoretical considerations \cite{Zavada:2009ska, Zavada:2011cv, DAlesio:2009cps}.

\begin{figure}[H]
\begin{center}
\includegraphics[scale=0.35]{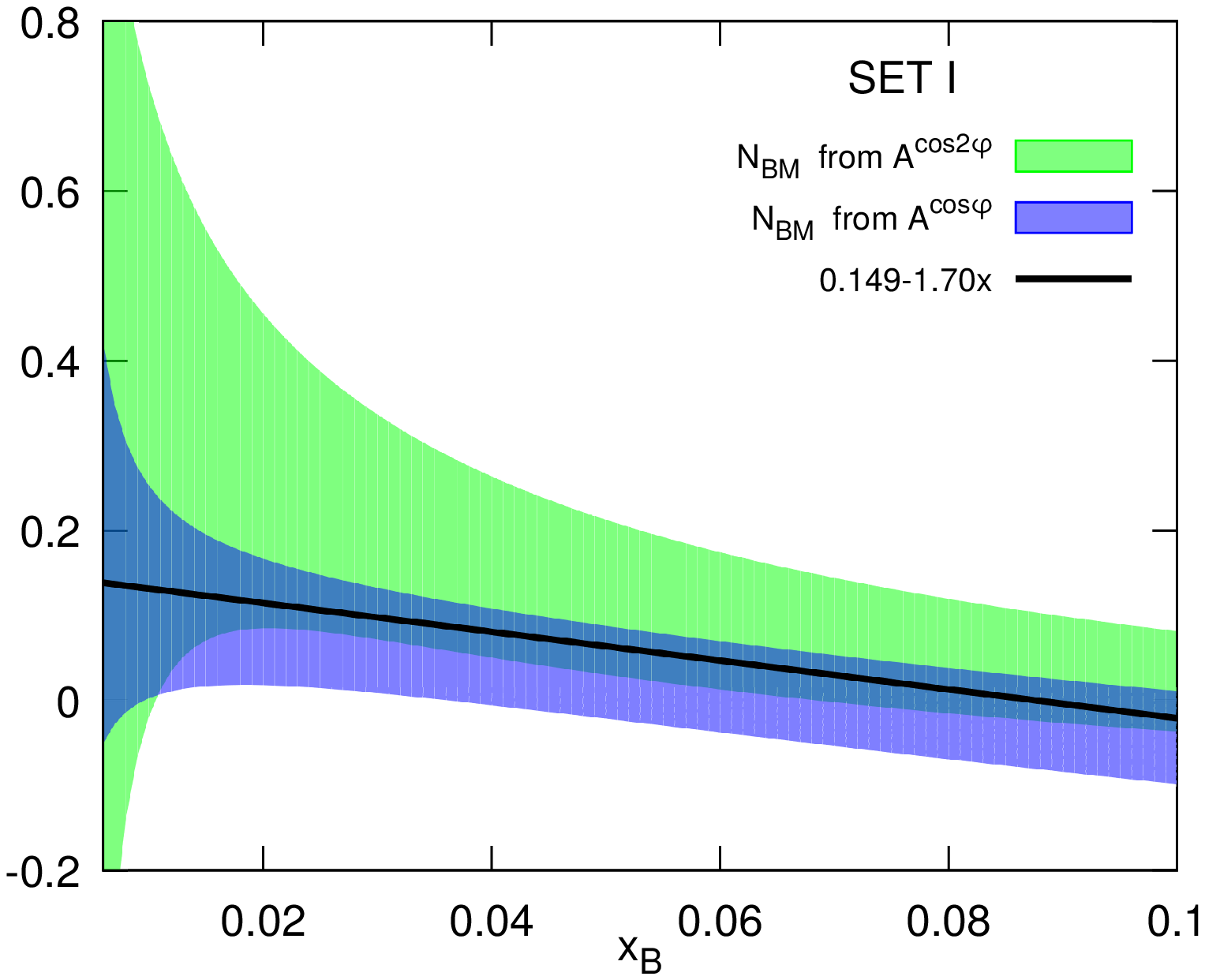}~~~\includegraphics[scale=0.35]{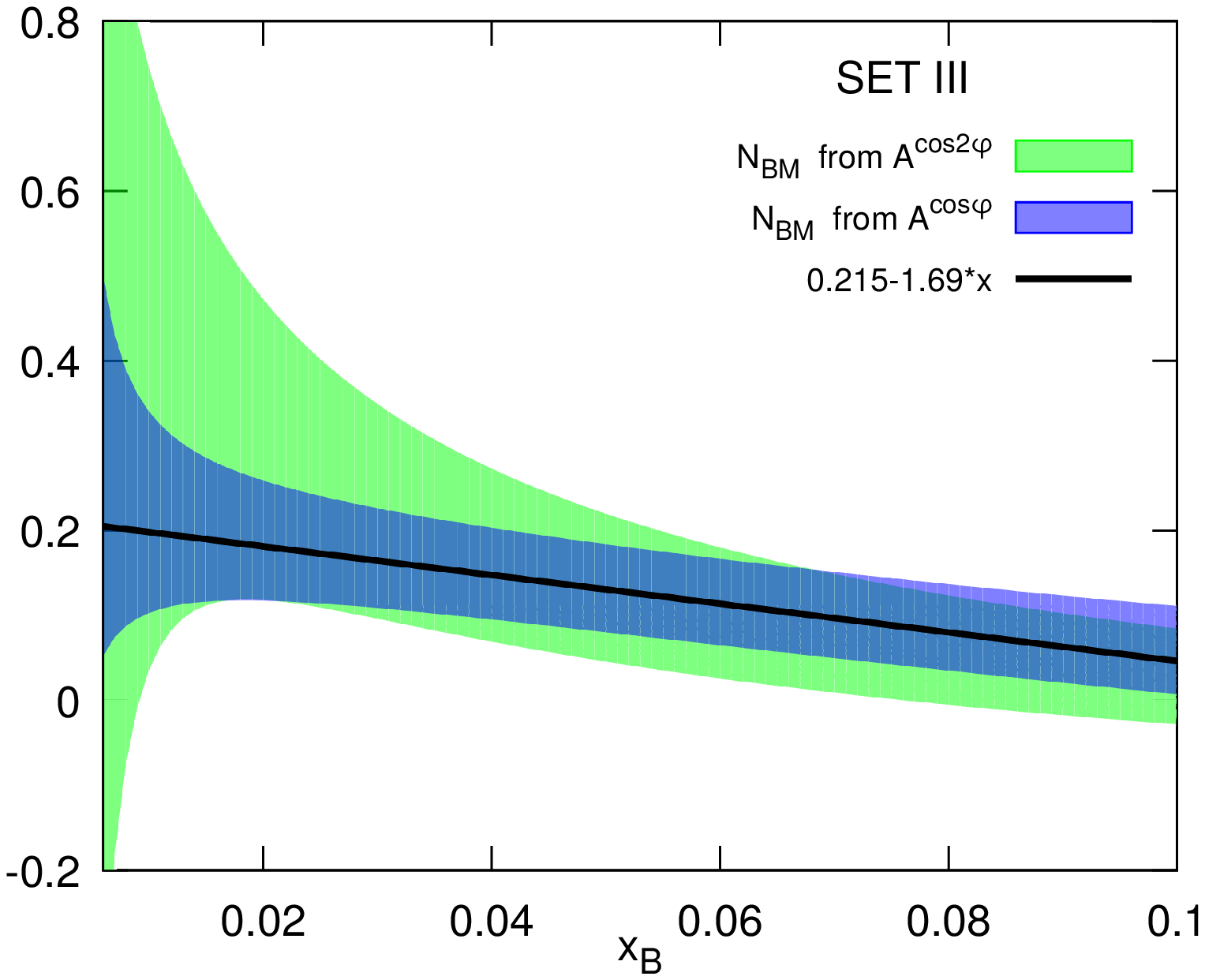}~~~\includegraphics[scale=0.35]{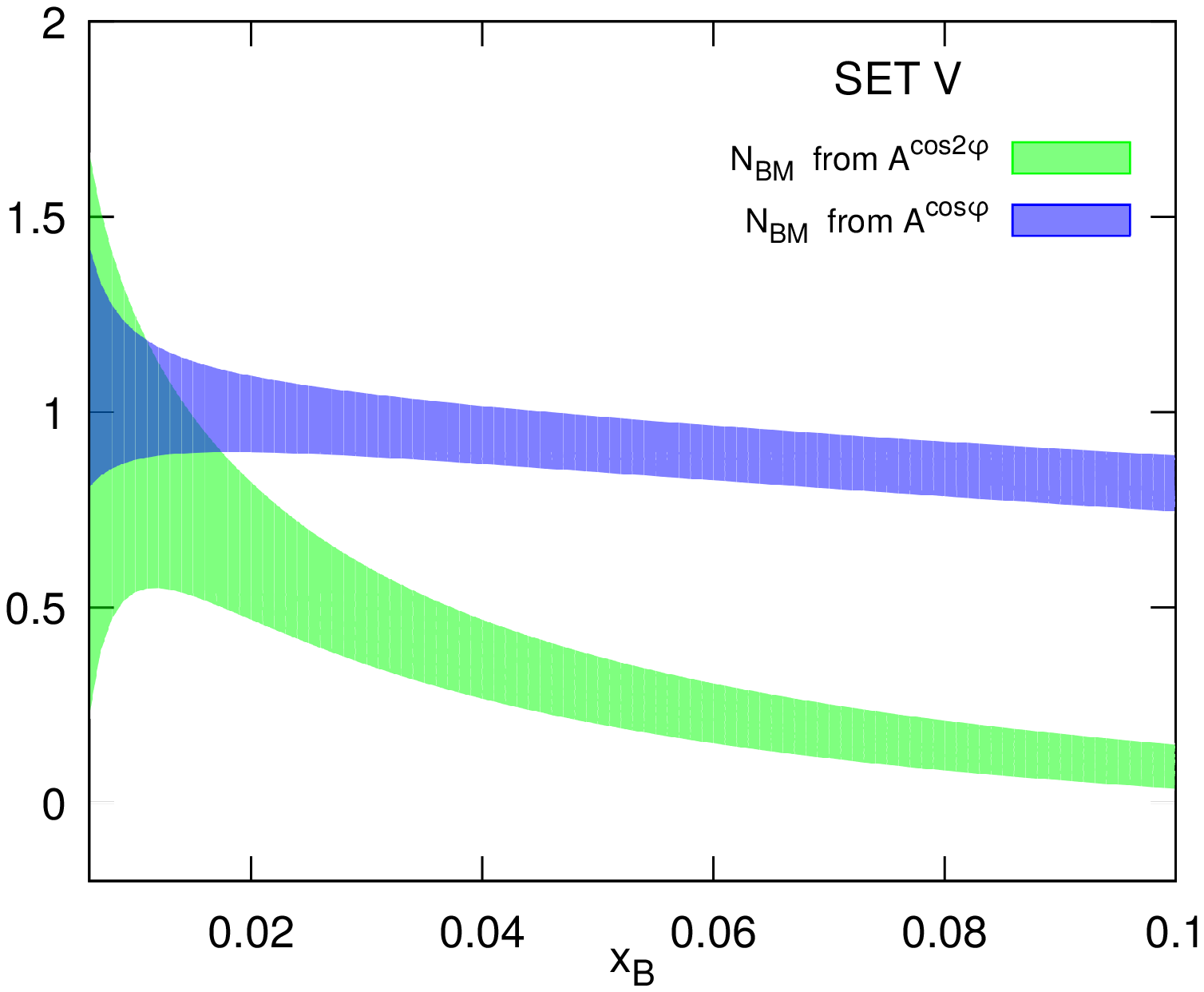}
\end{center}
\caption {${\cal N}\BM^{Q_V}(\xb)$ extracted from the difference asymmetries, Eqs.~(\ref{A1}) and (\ref{A2}), for
Sets I, III, and V. Plots for Sets II and IV overlap with those for Sets I and III, respectively.} \label{pic3}
\end{figure}

In Fig.~\ref{pic3}, we present our results  for Sets I, III, and V. The plots for Sets II and IV overlap with those for
Sets I and III, respectively, which implies that our analysis is not sensitive to  $M\BM$. Consequently, in the
following, we shall refer to Sets I, III, and V, only. \\
The excellent agreement with the data in Fig.3(b) suggests that the theoretical model, despite its simplifying assumptions,
gives a realistic description of the Boer-Mulders function in the kinematic regime of the COMPASS experiment.
\\

We obtain a simple linear fit to the extracted averaged ${\cal N}\BM^{Q_V}$ for the parameter Set~III, Eq. (\ref{best}): \be
{\cal N}\BM^{Q_V}(\xb ) &=& a\xb + b, \nonumber \\ a = -1.69\pm 0.08, \quad b &=& 0.215\pm 0.006. \ee

\subsection{Direct test for $\langle k_\perp ^2\rangle$ and $\langle p_\perp^2\rangle$}\label{Comp}


Interestingly, there is a second way to utilize equations (\ref{A1}) and (\ref{A2}) which  directly fixes the values of the
parameters $\avk ,\, \avp\,, M\BM$, and $M\C$ in Table~I. Eliminating ${\cal N}\BM^{Q_V}(\xb)$  from Eqs.~(\ref{A1}) and
(\ref{A2}) and using the variable $\rho$ we obtain, \be A(\xb )=B(\xb),\label{AB} \ee where
 \be
 A(\xb) &\equiv& \sqrt\frac{\langle Q^2(x_B)\rangle}{\avk}\,A_{UU,d}^{\cos\phi_h,h^+ -h^-}(\xb)\, +
\rho\,A_{UU,d}^{\cos2\phi_h,h^+ -h^-}(\xb),\\
 B(\xb) &\equiv& {\cal C}_{Cahn} + \rho\, \frac{\avk}{\langle Q^2(x_B)\rangle}\,\hat {\cal C}_{Cahn}
 \ee
 and the explicit expression for $\rho$ is
 \be
 \rho &=&\,- \frac{{\cal C}\BM}{\hat {\cal C}\BM}\,\nn
 &=&2\,\frac{\int dz_h\,[\Delta^N D_{{q_V}\!\uparrow}^{h}(z_h)]\,
[\,z_h^2\lambda\BM\,S_3(a,b,\avPT\BM)+\,(\eta\lambda\C-z_h^2\lambda\BM)\,S_1(a,b;\avPT\BM)]/(z_h^2\lambda\BM+\eta\lambda\C)^{3/2}}{\int
dz_h\,[z_h\,\Delta^N D_{{q_V}\!\uparrow}^{h}(z_h)]/(z_h^2\lambda\BM+\eta\lambda\C)\,S_2(a,b;\avPT\BM)}\label{rho} \ee

\begin{figure}[H]
\begin{center}
\includegraphics[scale=0.4]{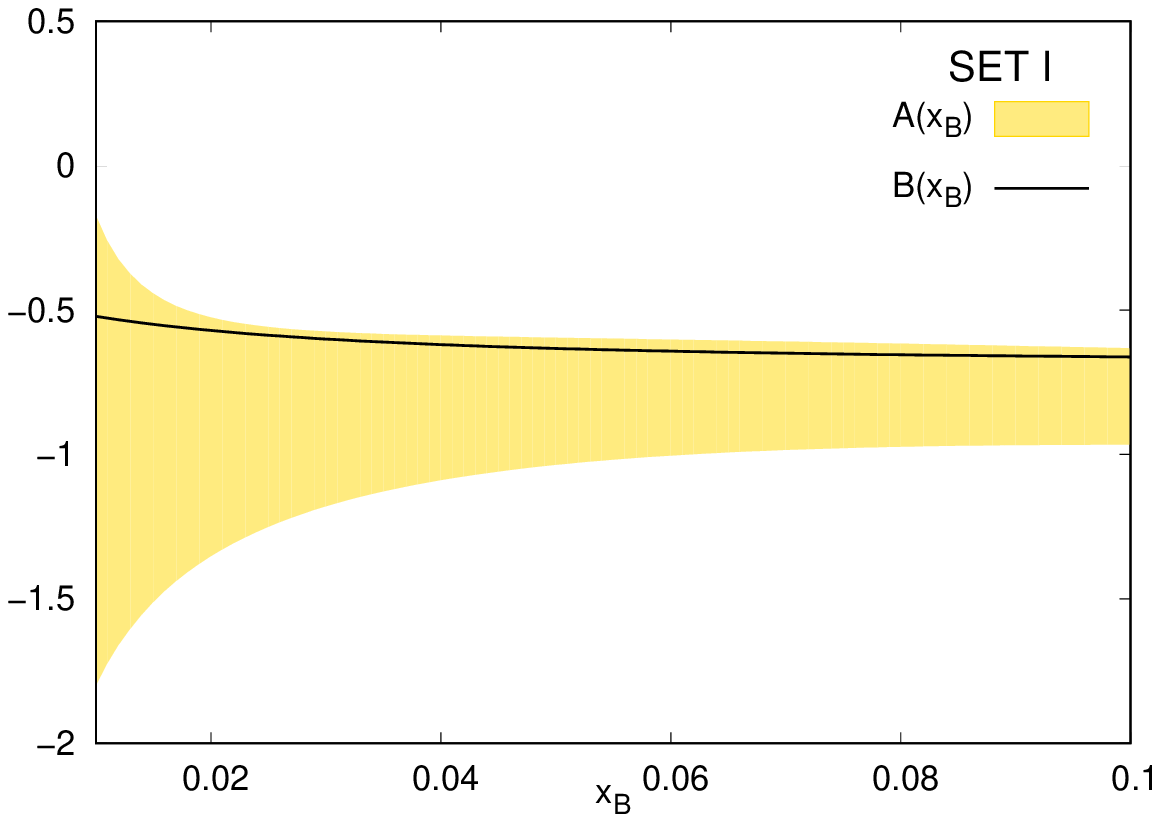}~~~\includegraphics[scale=0.4]{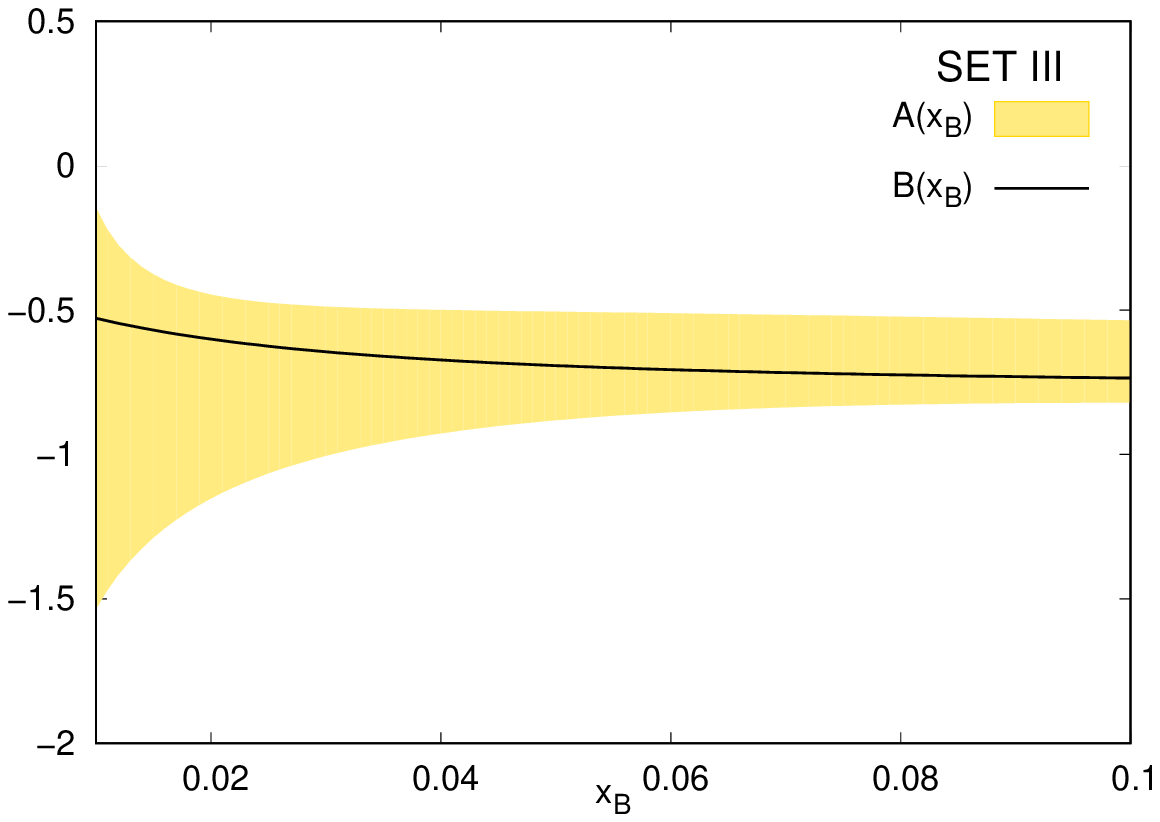}~~~\includegraphics[scale=0.4]{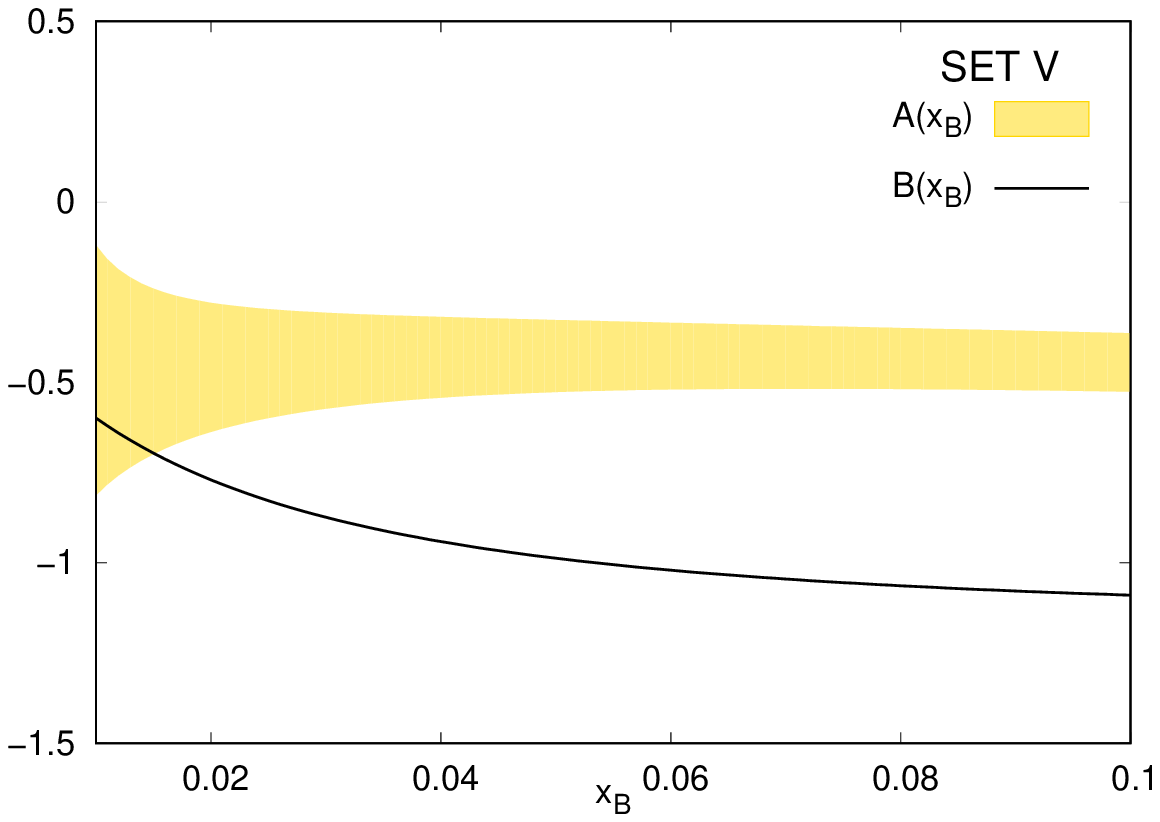}
\end{center}
\caption {The test of Eq.~(\ref{AB}) for Sets I, III, and V. Plots for Sets II  and IV  overlap with those for Sets I
($\chi^2_{d.o.f}=0.572$) and III ($\chi^2_{d.o.f}=0.151$), respectively.} \label{pic4}
\end{figure}

Figure~\ref{pic4} compares  the two functions $A(\xb )$ and $B(\xb )$ for Sets I, III, and V.\\

One  sees from  Fig.~\ref{pic4} that the COMPASS data on $A_{UU}^{\cos  \phi_h}$ and $A_{UU}^{\cos  2\phi_h}$, while roughly
compatible with  Set I again  favour  the parameter values of Set~III, Eq. (\ref{best}).\nl

(Note that for calculating $\chi^2$ in Fig. \ref{pic4} we use the COMPASS data points, \be \chi^2=\sum_{i=1}^N \frac{\left(
A(x_i)-B(x_i)\right)^2}{\left(\Delta A(x_i)\right)^2} \ee divided by the degrees of freedom d.o.f= $N-l$, $l$ is the number of
free parameters in the fit, in this fit $l=0$. Here $A(x_i)$ and $B(x_i)$ are values of the experimental points, $\Delta A(x_i)$
are the errors at $x_i$ calculated from Eq. (\ref{diff}). In this way, we obtain $\chi^2$ for each of five sets in Table~I
testing which of them fits the data the best.)
The compatibility of the two sides of Eq.~(\ref{AB}) constitutes a further test of the  simplifying assumptions made
in our analysis.

\section{A word of caution: evolution, interaction terms, and higher twist}
  In the analysis above, as mentioned several times,  we have not attempted to take into account any evolution in $Q^2$.
  As we shall explain, there are several reasons for this.
  
  The mechanism of TMD evolution is formulated in terms of functions  $\tilde{f}(x_B,\bm{b}_T, Q^2)$ where $\bm{b}_T$ is the
  Fourier transform variable conjugate to $\bm{k}_T$,

 \beq f(x_B,\bm{k}_T, Q^2) = \int \, d^2 \,\bm{b}_T \, e^{i\bm{k}_T\cdot \bm{b}_T} \, \tilde{f}(x_B, \bm{b}_T, Q^2) \eeq

 The evolution between two values $Q_0$ and $Q$ is mainly controlled by a factor

  \beq 
  \textrm{exp}\left\{\textrm{ln} \left(\frac{Q}{Q_0}\right) \,\tilde{K}(\bm{b}_T, Q_0) \right\} 
  \eeq
 where the Collins kernel $\tilde{K}(\bm{b}_T, Q_0)$ can be evaluated \textit{perturbatively} only for \textit{small} values of
  $b_T$. It is therefore split into a perturbative piece and a function $g_K(b_T)$ representing the nonperturbative part, and
  which is determined from fitting experimental data.\nl

  It is generally agreed that at small $b_T, \, g_K(b_T)\propto b_T ^2$, modulo a slowly varying logarithmic factor, but the
  expression
  \beq g_K(b_T)= 1/2\, g_2 b_T^2 \eeq
  where $g_2$ is a parameter to be fixed from data, used in several papers for \textit{all} $b_T$, is certainly incorrect at
  large $b_T$. In fact, Collins and Rogers \cite{Collins:2014jpa} suggest that
  \beq g_K(b_T) \xrightarrow{b_T\rightarrow \infty} \textrm{constant}. \eeq

\begin{figure}[H]
\begin{center}
\includegraphics[scale=0.4]{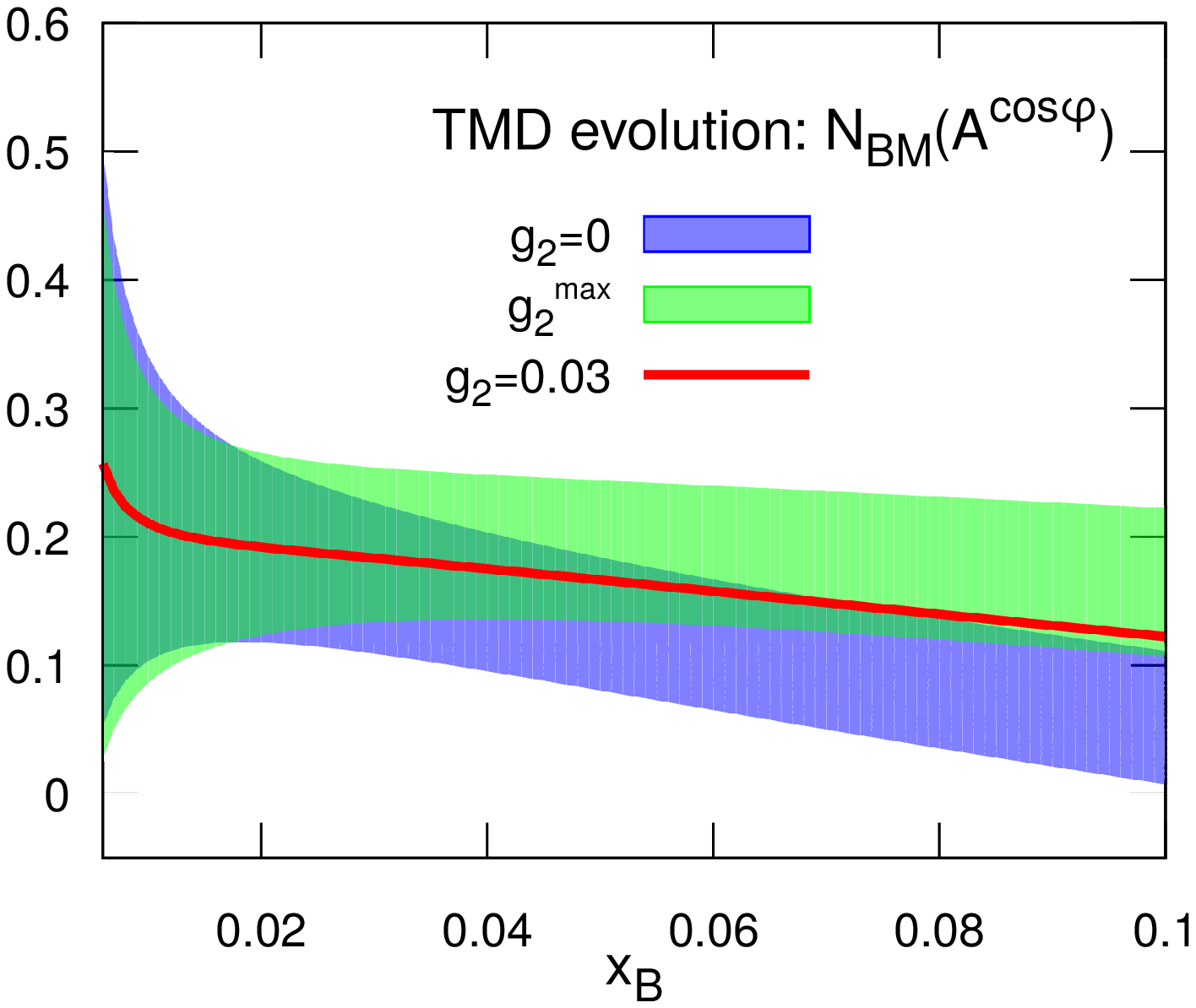}~~~~~~~\includegraphics[scale=0.4]{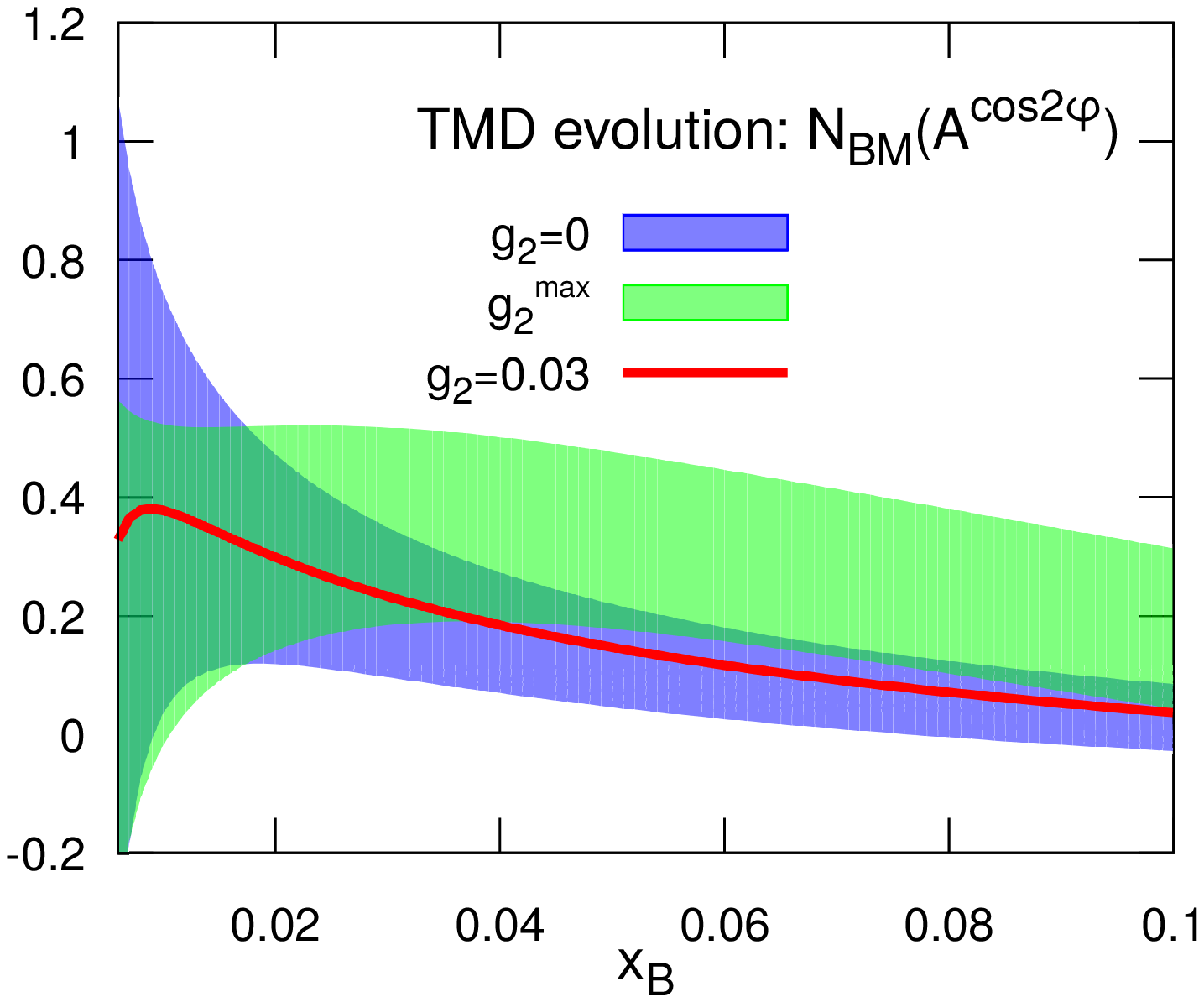}
\end{center}
\caption {TMD evolution effects for the function ${\cal N}\BM^{Q_V}(\xb)$ extracted from the difference asymmetries,
Eqs.~(\ref{A1}) and (\ref{A2}), for Set III (the same occurs also for other parameter sets).
We show results obtained using the two extreme values $g_2=0$ (no evolution) and $g_2=g_2^{\rm{max}}$ up to which the evolution
effects are negligible within experimental errors, the solid line shows ${\cal N}\BM^{Q_V}(\xb)$ for $g_2=0.03$,
given in \cite{Aidala:2014hva}.}
\label{pic5}
\end{figure}

  It should be clear that generally the bigger the range of $Q^2$ covered   by the data being fitted, the more accurate will be
  the determination of the function $g_K(b_T)$. Consequently, the values of $g_2$ reported in the literature  are heavily
  influenced by DY reactions and $Z$ production. Also, the shapes of the experimental distributions at large $Q^2$ suggest that
  the greatest sensitivity is to \textit{small} values of $b_T$ and hence the extracted large $b_T$ behaviour could be quite
  misleading for the much lower $Q^2$ SIDIS reactions. A wide range of values for $g_2$ are given in the literature, varying
  from 0 to  0.90. \nl
  As an example, in an exploratory study, Anselmino {\it et al.} \cite{Anselmino:2012aa} used the value $g_2 = 0.68 $. But
  this implies that if a parton density has a width $\avk\approx 0.20 \,{\rm GeV^2}$ at $Q_0^2=1 \,{\rm GeV^2}$,
  then at $Q^2=2.4  \,{\rm GeV^2}$ the width has grown to $\avk\approx 0.80 \,{\rm GeV^2}$,  surely a totally unphysical increase. \nl
  And in a later study focused on the SIDIS $Q^2$ range, Aidala {\it et al.} \cite{Aidala:2014hva} suggested that
  $0\, \leq g_2 \,\leq\, 0.03 $.
  The limited amount of data and the small range of $Q^2$ involved 
  suggest that these measurements  are not the most suitable for studying the structure of TMD evolution and we carried out our analysis ignoring TMD
  evolution. However, to give some feeling for possible evolution effects,  in Fig.~\ref{pic5} we plot the function
  ${\cal N}\BM^{Q_V}(\xb)$ taking into account evolution. We use the two extreme values $g_2=0$ (no evolution)
  and $g_2^{\rm{max}}$ up to which the evolution effects are negligible within experimental errors.
  We also present the results for $g_2=0.03$, given in \cite{Aidala:2014hva}. 
  We find $g_2^{\rm{max}}=0.05$ for the asymmetry $A^{\cos\phi}_{UU}$ and $g_2^{\rm{max}}=0.35$ for $A^{\cos 2\phi}_{UU}$.
  
  Thus, in our analysis, we can neglect the evolution up to $g_2=0.05$, which is in agreement with \cite{Aidala:2014hva},
  keeping in mind all above comments.   

The expression Eq.~(\ref{A1}), which  we have used above for the asymmetry $A_{UU}^{\cos  \phi_h, h-\bar h}(\xb)$, is
incomplete. There are so called interaction-dependent terms \cite{Bacchetta:2006tn}, linked to the quark-gluon-quark
correlators, which  have been left out. These interaction terms are unknown, but  we might expect to have roughly
 \beq |\textrm{Interaction-dependent terms}| \sim \alpha_s |\textrm{terms in }\textrm{Eq.}~(\ref{A1})|. \eeq
In our kinematic range, the average value of $\alpha_s$ is approximately $0.35$, which suggests that such terms might
be small compared to the terms kept in Eq.~(\ref{A1}).
Attempts have been made in the literature to estimate the size of terms of this type, at least where they occur in the
difference between the true value of the structure function $g_T$ and  the Wandzura-Wilczek (WW) approximation $g_T^{WW}$
to it. The  data on $g_T$ were compared  with $g_T^{WW}$ by Accardi {\it et al.} \cite{Accardi:2009au}, who claimed differences
of order $15\%-35\%$.
However, the data are of very poor quality so that the conclusion reached in \cite{Accardi:2009au} does  not seem convincing.
On the other hand, a recent lattice calculation of $g_T$ by Bhattacharya {\it et al.} \cite{Bhattacharya:2020cen} found very good
agreement with $g_T^{WW}$  out to $x_B=0.4$. In any event, the terms neglected in the WW approximation to $g_T$ are not
the same as those ignored in the $\cos \phi_h$ asymmetry, so these results can only be considered as a hint that the terms
ignored in the $\cos \phi_h$ asymmetry are indeed negligible. \nl
It should be noted that in their recent general study of asymmetries in SIDIS, in the section on $\cos 2\phi_h$ and
$\cos \phi_h$ asymmetries, Bastami {\it et al.} \cite{Bastami:2018xqd} found significant differences between
the $\cos \phi_h$ asymmetry data and what they refer to as the WW approximation to it.
However, Bastami {\it et al.} \cite{Bastami:2018xqd} utilized the results of \cite{Barone:2009hw, Barone:2010gk},
which, as the present study demonstrates, are incorrect.

Ultimately, the excellent fit to the data found in this study, together with the successful consistency test,
suggests that the omission of these terms is justified. \nl

\begin{figure}[H]
\begin{center}
\includegraphics[scale=0.4]{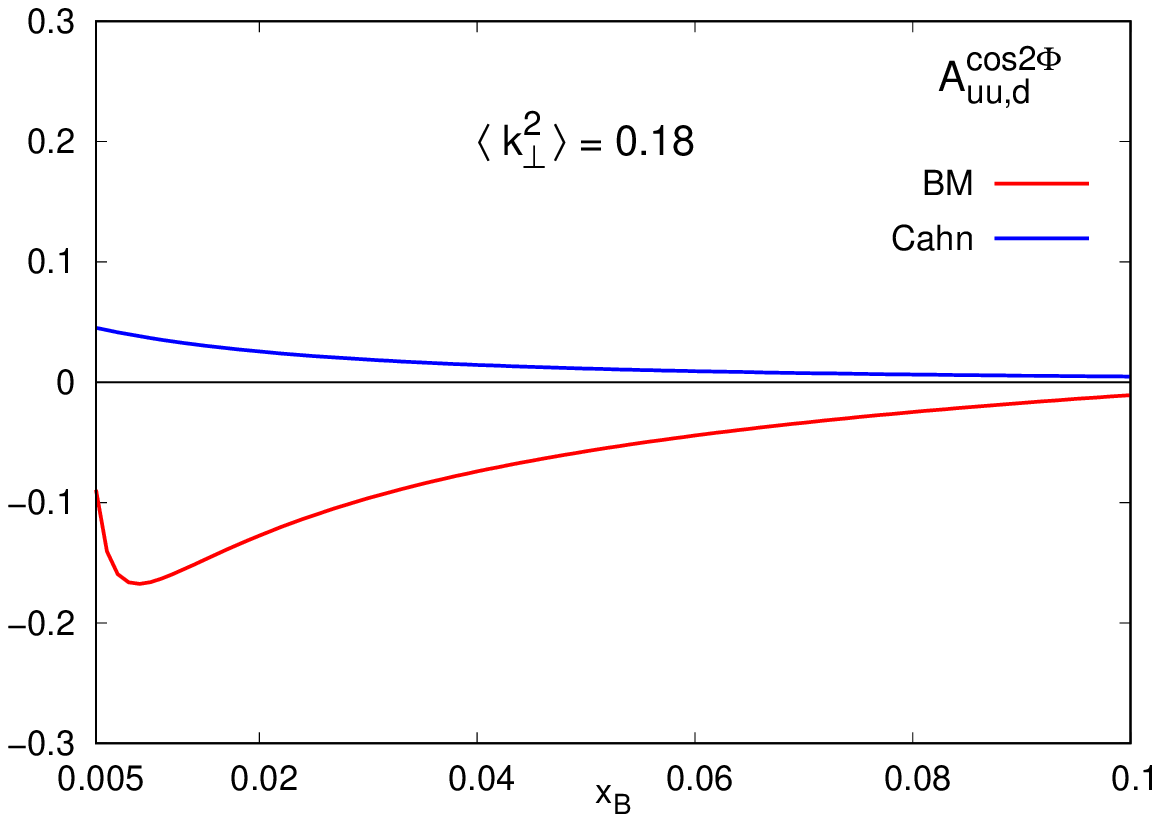}~~~\includegraphics[scale=0.4]{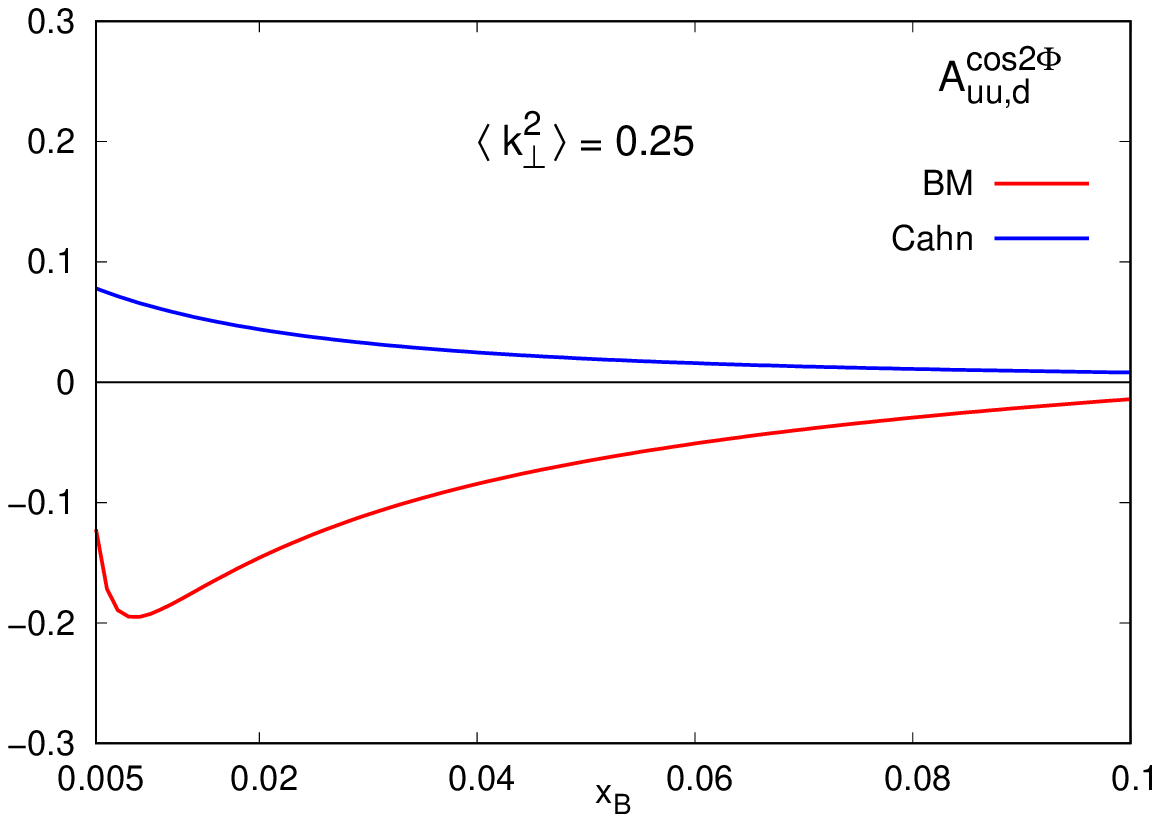}~~~\includegraphics[scale=0.4]{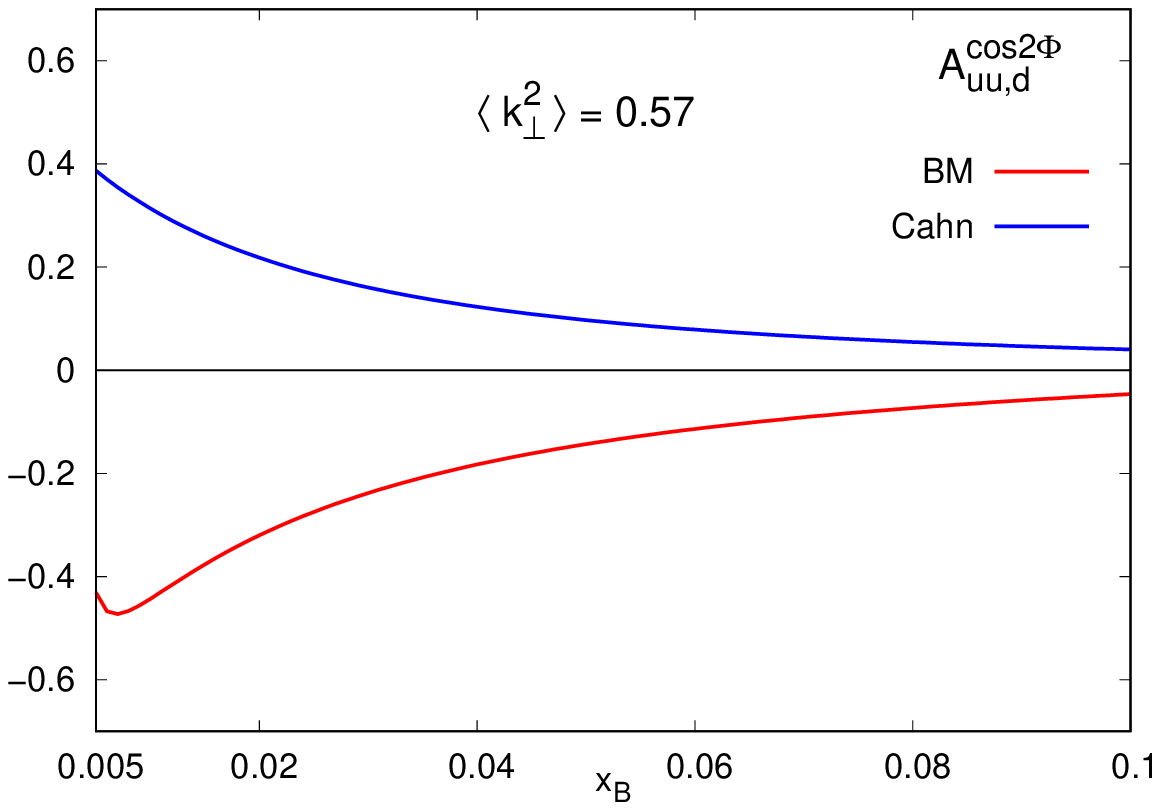}\\
\end{center}
\caption {Comparison of the BM with Cahn contribution to the $\cos 2\phi_h$ asymmetry for $\avk=0.18$ (left), $\avk=0.25$
(middle), and $\avk=0.57\,{\rm GeV^2}$ (right).} \label{pic7}
\end{figure}

Finally, it should be noted that the expression Eq.~(\ref{A2}) for the asymmetry  $A_{UU}^{\cos  2\phi_h, h-\bar h}(\xb )$
is unusual in that it contains a combination of a twist-2 BM  term  ${\cal N}\BM^{Q_V}(\xb)\,
   \hat {\cal C}\BM^h$    with a twist-4 Cahn
 term $\avk/\langle Q^2\rangle\,\cdot\,\hat {\cal C}^h_{Cahn}$, and one might wonder whether there might exist important
 twist-4 BM terms  which are not accounted for in Eq.~(\ref{A2}). That this is not so can be understood  from the following
 argument.
 In Fig.~\ref{pic7}, we compare the BM and Cahn contributions to Eq.~(\ref{A2}).
 Remarkably, the twist-4 Cahn term contribution is not negligibly small in magnitude compared to the twist-2 BM contribution!
 This peculiar situation is due to two factors. First,
 the twist-4 prefactor in the Cahn term,
 $\avk/\langle Q^2\rangle$, is not really small for the  values of  $Q^2$ in our data. Second, the Cahn factor
 $\hat {\cal C}^h_{Cahn}$ is anomalously large because it depends on the \textit{unpolarized } PDFs and FFs.
 This suggests that any further twist-4 BM  type
 contribution would  be expected to be negligibly small by comparison and it also explains our earlier comment that the
 neglect of this Cahn term in  \cite{Barone:2015ksa} is dangerous and unjustified. \nl

\section{Comparison to other Boer-Mulders parametrizations}

\begin{figure}[H]
\begin{center}
\includegraphics[scale=0.35]{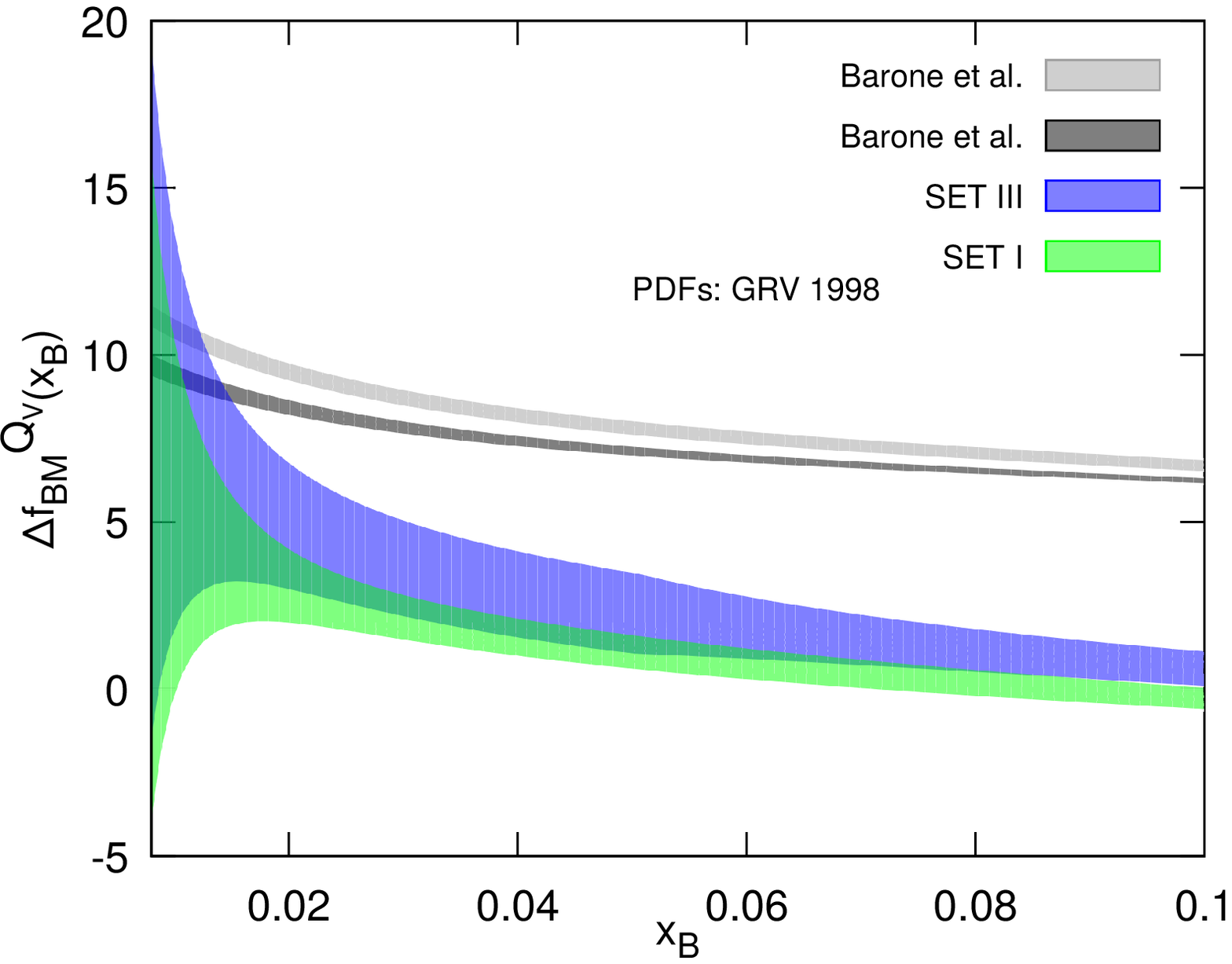}~~~\includegraphics[scale=0.35]{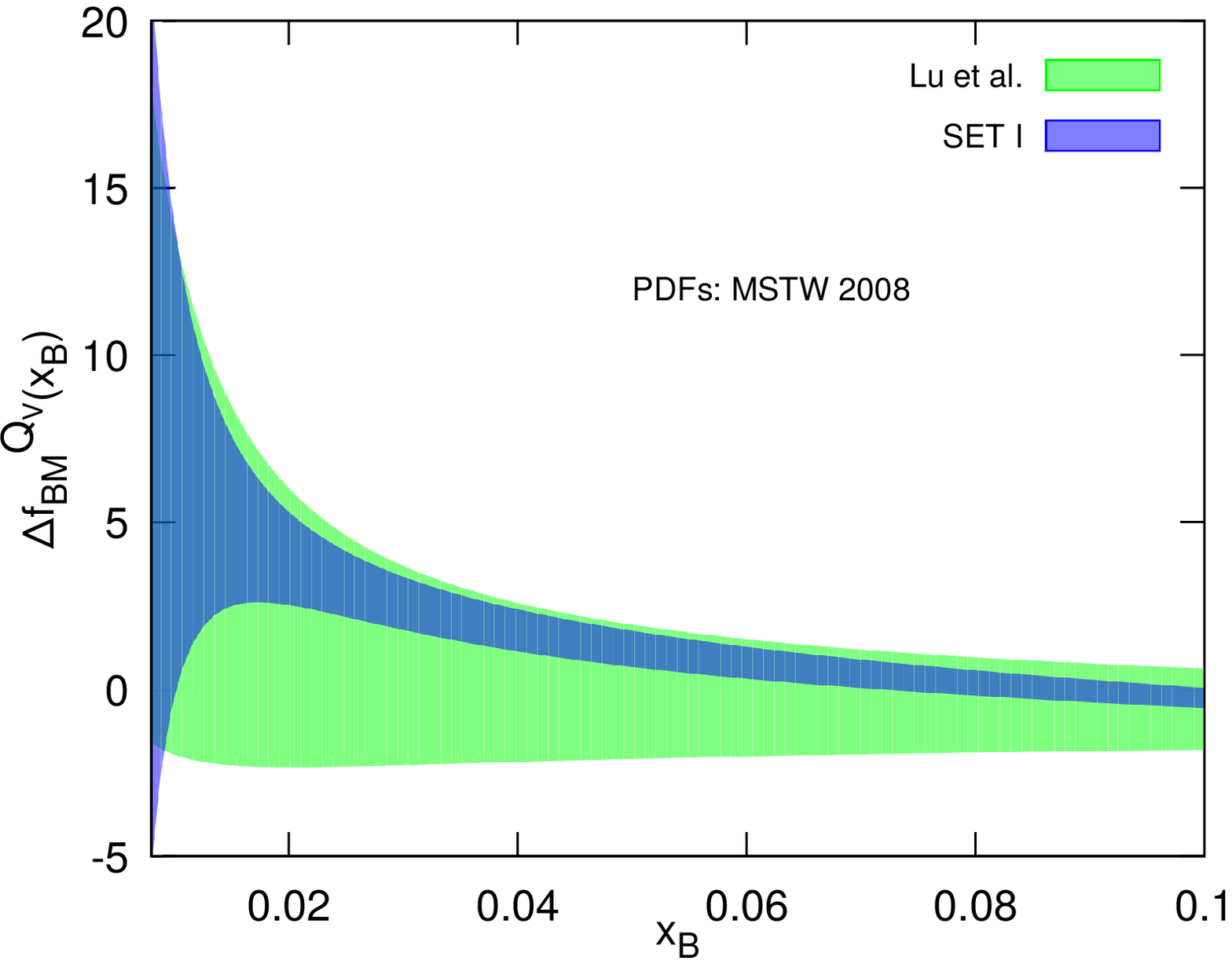}~~~\includegraphics[scale=0.35]{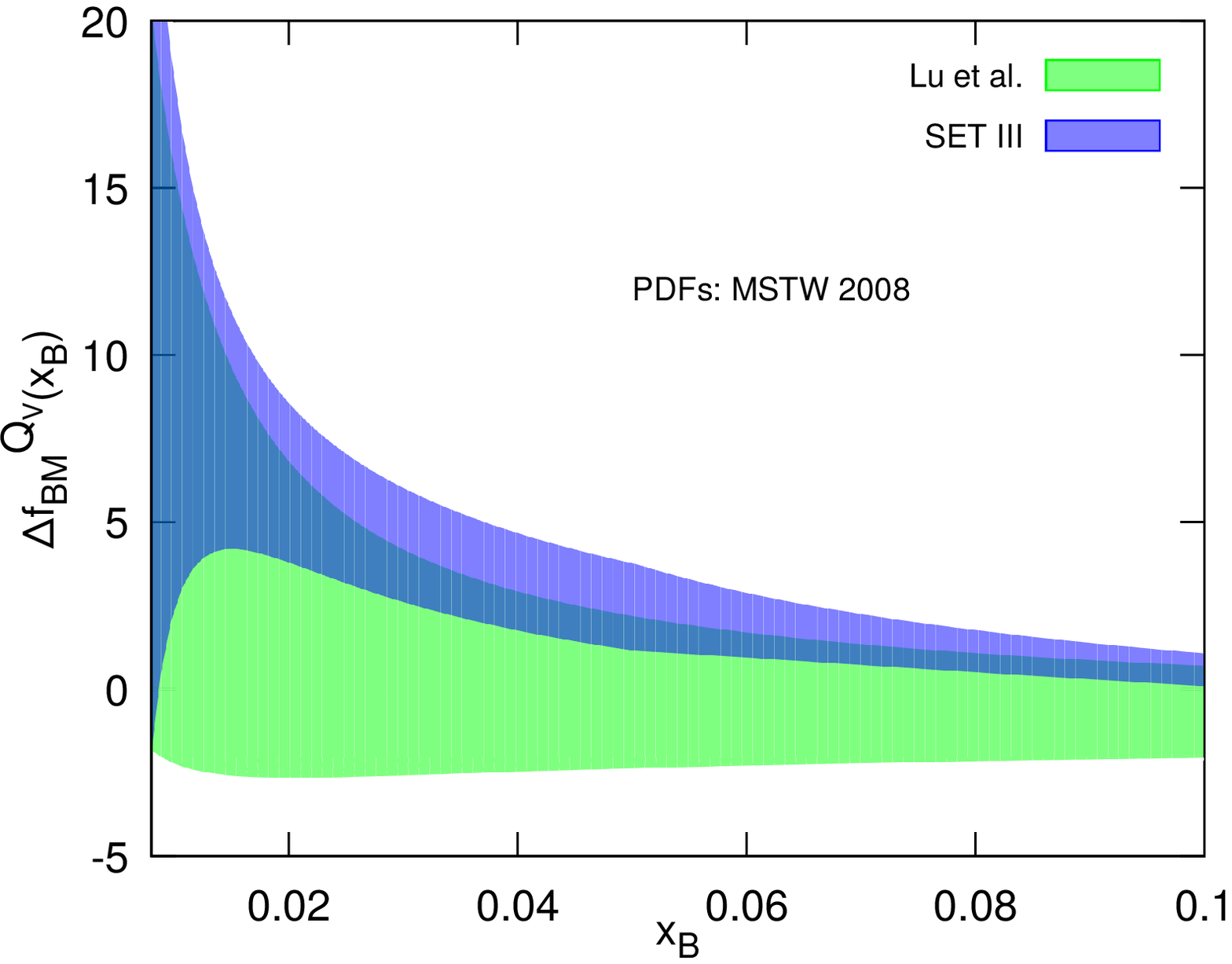}
\end{center}
\caption{Comparison of $\Delta f\BM^{Q_V}$ for  Sets I and III with the result of Barone {\it et al.} (left) \cite{Barone:2009hw}
(light gray) and \cite{Barone:2010gk} (dark gray) and
with Lu and Schmidt \cite{Lu:2009ip}--middle (Set I) and right (Set III). We
use, respectively, GRV1998 \cite{Gluck:1998xa}
and MSTW2008 \cite{Martin:2009iq} parametrizations for the collinear PDFs.} \label{pic8}
\end{figure}

Our valence Boer-Mulders function $\Delta f\BM^{Q_V}(\xb)$, \beq \Delta f^{Q_V}\BM (\xb,Q^2)\!=\! 2\,{\cal N}\BM
^{Q_V}(\xb)\,Q_V(\xb,Q^2)\, , \label{fbm} \eeq is shown in Fig.~\ref{pic8}, where it is compared to $\Delta f\BM^{Q_V}(\xb)$
calculated using two other parametrizations of BM functions available in literature--the BM functions published in
\cite{Barone:2009hw, Barone:2010gk} and in \cite{Zhang:2008nu, Lu:2009ip}. The BM function published in \cite{Barone:2009hw,
Barone:2010gk} is extracted from the $\cos 2\phi_h $ asymmetry in SIDIS, using  the simplifying, but theoretically inconsistent,
assumption that it is proportional to the Sivers function for {\it each quark flavour separately}. The parametrizations in
\cite{Zhang:2008nu, Lu:2009ip} are extracted from the azimuthal  $\cos 2\phi_h $ asymmetry of the final lepton pair in unpolarized
Drell-Yan processes. We compare our result to the parametrization  in \cite{ Lu:2009ip}, obtained from the combined analysis of
the $pp$ and $pd$ DY  processes.\\

It is seen that, both for Sets I and III, there is a significant difference between our predictions and those of
Refs. \cite{Barone:2009hw, Barone:2010gk}, and a good agreement with the results in \cite{Lu:2009ip} from DY data.
This suggests that the BM functions in \cite{Barone:2009hw, Barone:2010gk} are incorrect.

\section{Test of the Boer-Mulders to Sivers relation}

In Refs. \cite{Barone:2008tn, Barone:2009hw, Barone:2010gk}, the BM functions were assumed proportional to the Sivers functions
for each quark and antiquark flavour $q$ separately, \beq \Delta f^q \BM (\xb ,\kt )= \lambda_q\, \Delta f^q_{Siv}(\xb ,\kt ),
\label{te}
 \eeq
which, as implied by the results in Fig.~\ref{pic8} above, is badly violated, in agreement with conclusions reached in our
earlier paper \cite{Christova:2015jsa}. We here  return to the question of the proportionality between the BM and Sivers
functions, but now \emph{only} for the valence-quark contributions $Q_V=u_V+d_V$, \beq \Delta f^{Q_V} \BM (\xb ,\kt )=
\lambda\,\Delta f^{Q_V}_{Siv}(\xb ,\kt ) \label{Siv}
 \eeq
For the Sivers function we use an analogous  parametrization to the BM, Eq. (\ref{BM-Siv_dist1}), but with the replacements
$M\BM \to M\S$ and
 ${\cal N}\BM  \to {\cal N}_{Siv}$. Then, Eq. (\ref{Siv}) implies $M\BM =M\S$,  and
 \be
   {\cal N}\BM^{Q_V} (\xb ) = \lambda\,\, {\cal N}_{Siv}^{Q_V}(\xb )\label{testSiv}
  \ee
 which we shall now test.

We extract ${\cal N}_{Siv}(\xb )$ from the difference Sivers asymmetries $A_{UT}^{Siv,h^+-h^-}(\xb )$ using the single-spin
asymmetries presented by COMPASS for $h^\pm$ on deuterons \cite{Alekseev:2008aa}.
  The expression for $A_{UT}^{Siv,h^+-h^-}(\xb )$ is
  \cite{Christova:2015jsa}
 \be
 A_{UT,d}^{Siv,h^+-h^-}(\xb )&=&\sqrt{\frac{e\pi}{2}}
 \,\, K_{Siv}^h\,{\cal N}_{Siv}^{Q_V}(\xb ),\label{ASiv}
 \ee
 \be
  K_{Siv}^h =\frac{\avk\S^2}{M\S\avk }\,
  \frac{\int dz_h\,z_h[D_{q_V}^h]/\sqrt{\avPT\S}}{\int dz_h\,[D_{q_V}^h]},\quad
 \avPT\S =\avp +z_h^2 \avk\S
 \ee

\begin{figure}[H]
\begin{center}
\includegraphics[scale=0.5]{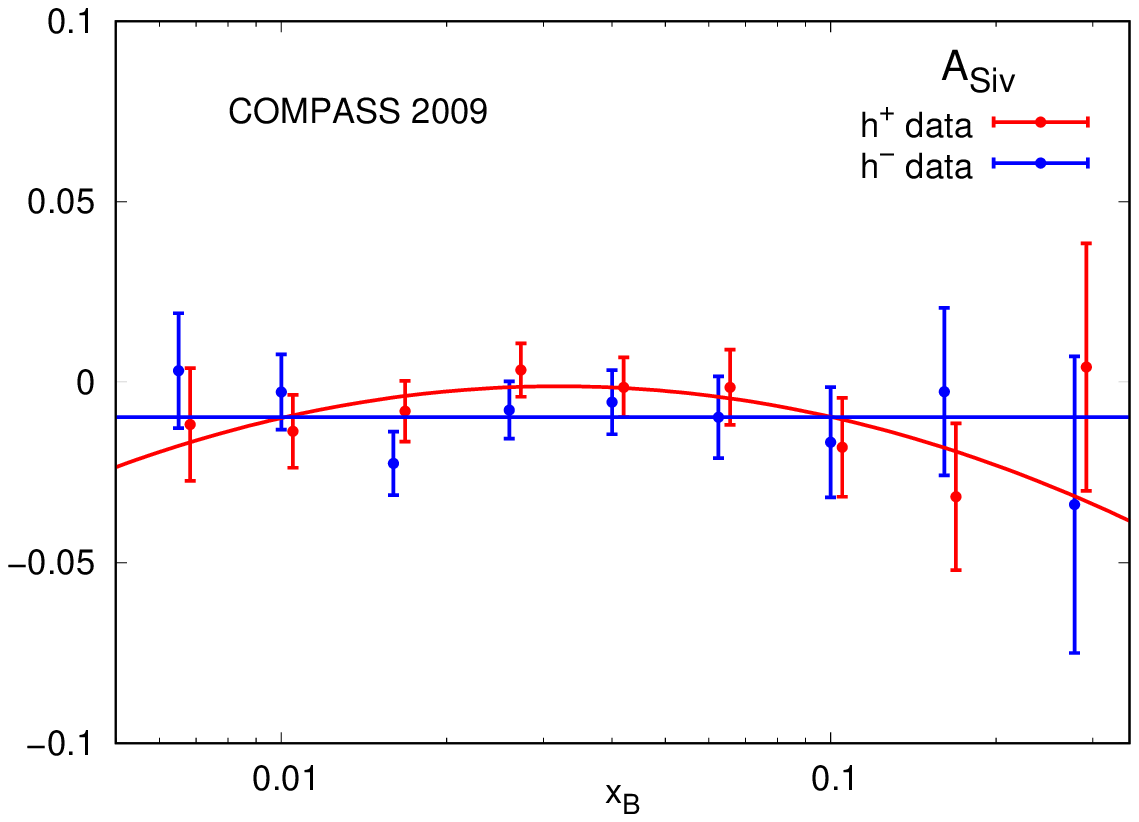}~~~\includegraphics[scale=0.5]{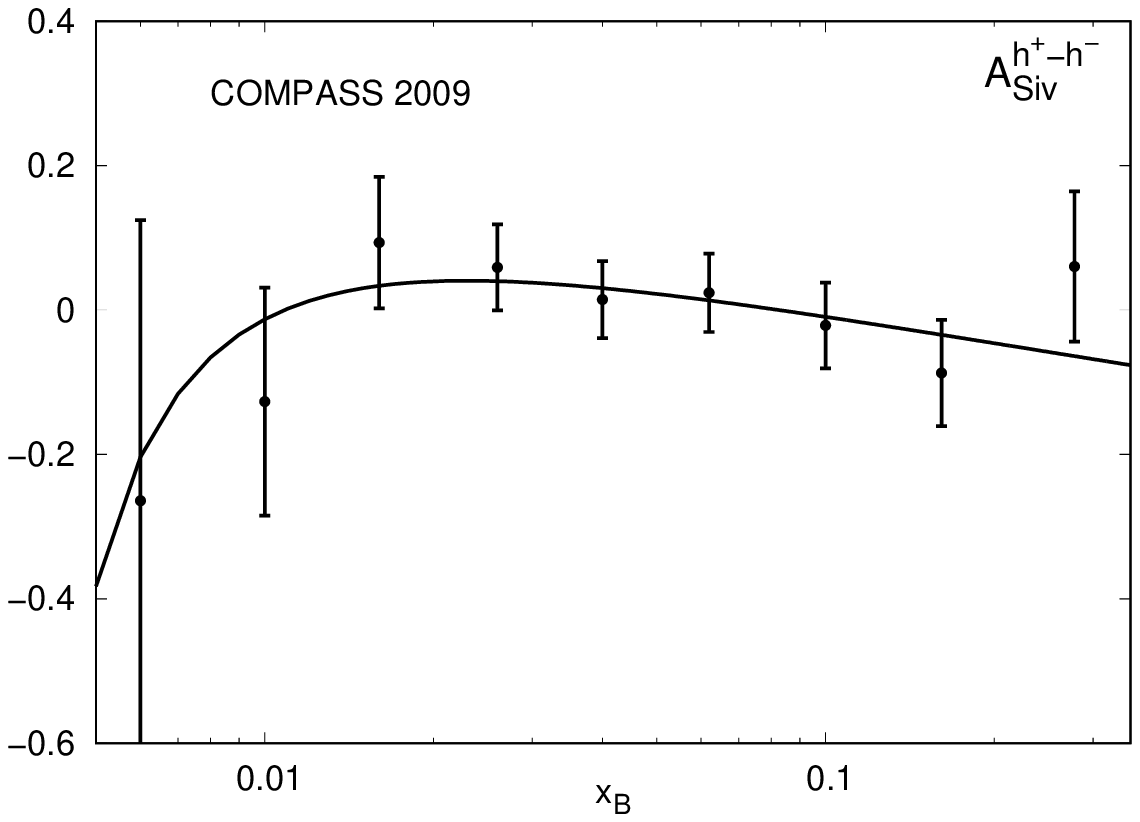}
\end{center}
\caption {Smooth fits to the Sivers asymmetries: $A_{Siv}^{h^\pm}$ (left) and $A_{Siv}^{h^+-h^-}$ (right).} \label{picSiv}
\end{figure}

\begin{figure}[H]
\begin{center}
\includegraphics[scale=0.4]{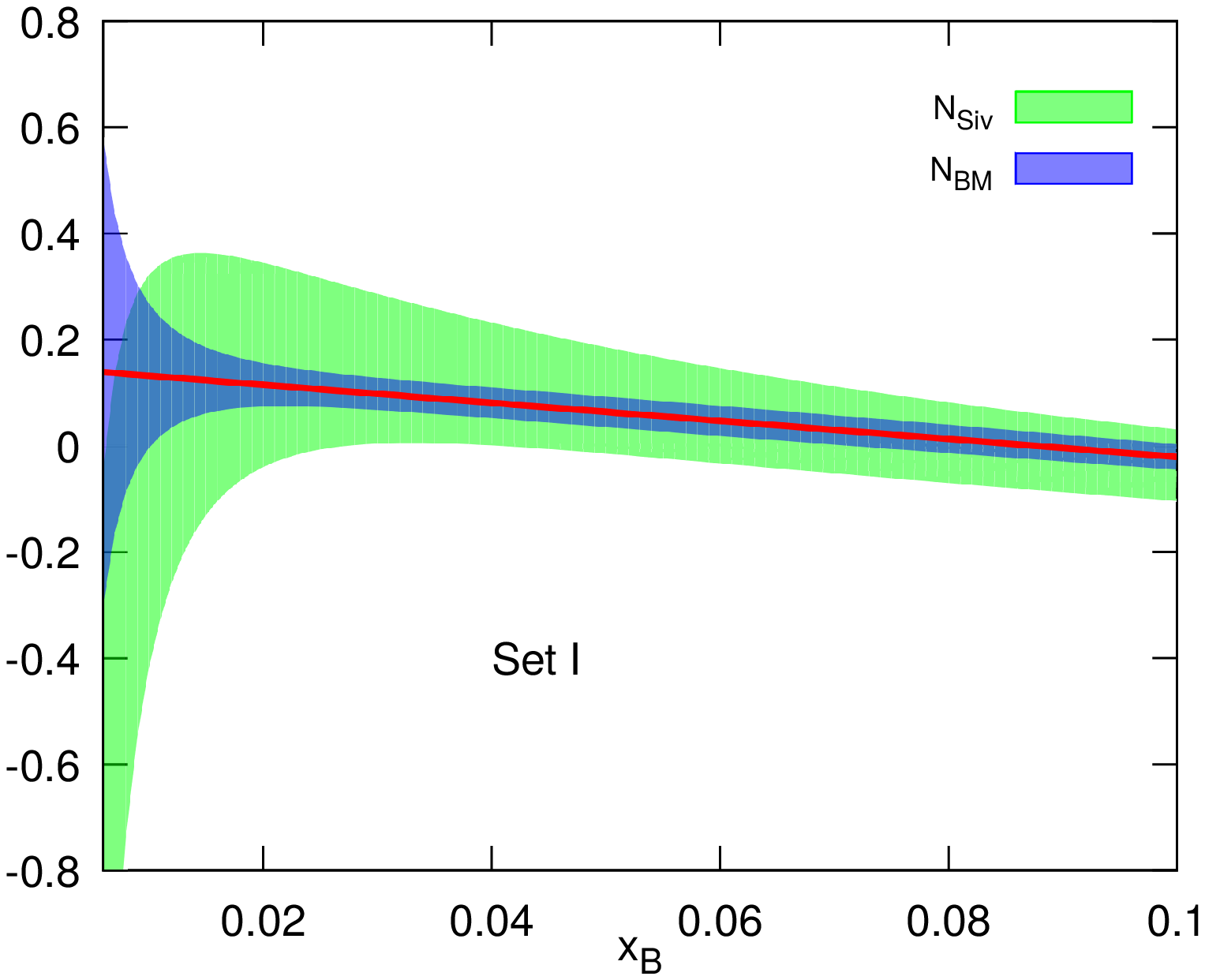}~~~\includegraphics[scale=0.4]{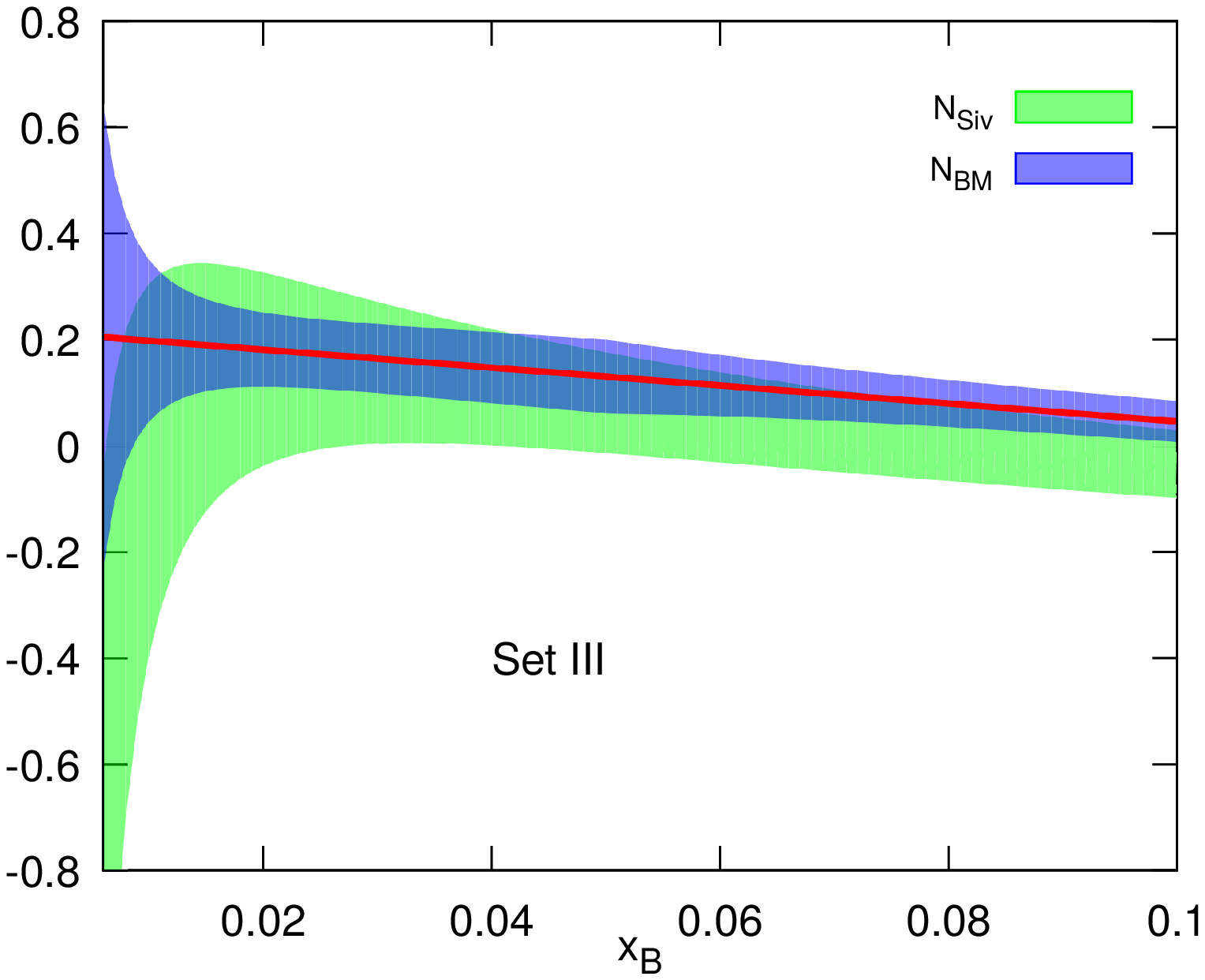}
\end{center}
\caption {Test of proportionality for BM to Sivers functions, Eq. (\ref{testSiv}). Both for  Set I  (left) and Set III (right),
${\cal N}\BM^{Q_V} (\xb )$
 overlaps   ${\cal N}_{Siv}^{Q_V}(\xb )$,  which implies  $\lambda \approx 1$ for both sets.}
\label{piccomp}
\end{figure}

 In Fig.\ref{picSiv}, we show the measured single-spin $A_{Siv}^{h^\pm}$ and difference $A_{Siv}^{h^+-h^-}$  Sivers asymmetries,
 and in Fig.\ref{piccomp}, we show the extracted BM and Sivers functions
 ${\cal N}\BM^{Q_V}(\xb )$ and ${\cal N}_{Siv}^{Q_V}(\xb )$ for Sets I and III.
  We see that Eq. (\ref{testSiv}) holds fairly well, confirming the results of \cite{Christova:2015jsa}, and that
  $\lambda\approx 1.0$ for both Set I  and  Set III.

\section{Conclusions}

 In a combined analysis of the $\cos\phi_h$ and $\cos 2 \phi_h$ azimuthal asymmetries in unpolarized SIDIS,
 measured most recently by COMPASS,  we determined (1) the BM function $f^{Q_V}\BM (\xb,Q^2)$ for the sum of the valence
 quarks $Q_V=u_V+d_V$ and (2) obtained information on
 the average transverse momenta $\avk$ and $\avp$, which play a role in the transverse momentum dependent
 parton distribution functions and fragmentation functions, respectively.
 The analysis is based on a study of the so-called difference asymmetries
 between hadron $h$ and $\bar{h}$. The results are obtained using the often made simplifying assumption  of
 factorization of transverse momentum and $x_B$ dependence, with the
 transverse momentum behaviour given by Gaussians, with  $\xb$-independent widths. The excellent agreement with the data, as
  well as the positive result of the compatibility test, Sec.~(\ref{Comp}), suggests that these simplifications are valid in the
  kinematic region of the COMPASS experiment.  \nl
  We have compared our results  to the existing ones in the literature.
  For the BM function, we agree with the results obtained from an analysis of DY processes \cite{Zhang:2008nu, Lu:2009ip} but
  disagree strongly  with the results  obtained in a model analysis of the $\cos 2\phi_h$ asymmetry in SIDIS in
  \cite{Barone:2009hw, Barone:2010gk}, both obtained under the same simplifying assumptions as in this paper.
  It should be noted that in their recent general study of asymmetries in SIDIS, in the section on $\cos 2\phi_h$ and
  $\cos \phi_h$ asymmetries, Bastami {\it et al.} \cite{Bastami:2018xqd} utilized the, according to our study,
   unreliable  results of \cite{Barone:2009hw, Barone:2010gk}. \nl
   Our favoured values for $\avk$ and $\avp$ agree with those obtained in a previous analysis of the $\cos\phi_h$
   and $\cos 2 \phi_h$ modulations in SIDIS, \cite{Anselmino:2005nn, Giordano:2008th},
   i.e., ($\avk =0.18\;{\rm GeV^2};\avp =0.20\;{\rm GeV^2}$) and ($\avk =0.25\;{\rm GeV^2};\avp =0.20\;{\rm GeV^2}$), and
   disagree with the later, larger values ($\avk =0.57\,{\rm GeV^2};\avp =0.12\,{\rm GeV^2}$) obtained from a study
    of multiplicities \cite{Anselmino:2013lza}. \nl
Finally, we note that  future data on the $\langle\cos\phi_h\rangle$ and $\langle\cos 2\phi_h\rangle$ asymmetries on protons, for
charged pions or kaons, will allow access to the BM function for the valence quarks $u_V$ and $d_V$ separately, in the same,
approximately model-independent manner \cite{Christova:2015jsa}.\\

\section{Acknowledgements}
E. C. and D. K. acknowledge the support of the INRNE-BAS (Bulgaria)-JINR (Russia) collaborative grant.
E. C. is grateful to Grant No. 08-17/2016 of the Bulgarian Science Foundation and 
D. K. acknowledges the support of the Bogoliubov-Infeld Program.
D. K. also thanks A. Kotlorz for useful comments on numerical analysis.

\end{document}